\documentclass[letterpaper,twocolumn,10pt]{article}
\usepackage{usenix-2020-09}

\usepackage{xspace}
\usepackage{booktabs}
\usepackage{graphicx}
\usepackage{amsmath}
\usepackage[ruled,vlined, noend]{algorithm2e}
\usepackage{algorithmic}
\usepackage{enumitem}
\usepackage{scalerel}
\usepackage{array, caption, tabularx,  ragged2e,  booktabs}

\def\BibTeX{{\rm B\kern-.05em{\sc i\kern-.025em b}\kern-.08em
    T\kern-.1667em\lower.7ex\hbox{E}\kern-.125emX}}

\newcommand{\acltestcap}{ACtest\xspace}
\newcommand{\acltest}{ACtest\xspace}

\newcommand{\acltestgen}{ACtestGen\xspace}

\newcommand{\numoftestedversions}{193}
\newcommand{\numofvulversions}{72}
\newcommand{\numofdetectedvul}{168}
\newcommand{\numofreportedvul}{135}

\newcommand{\sysnet}{Center\xspace}
\newcommand{\opera}{Group\xspace}
\newcommand{\ucsdcse}{Course\xspace}
\newcommand{\company}{Company\xspace}

\usepackage{sectsty}
\sectionfont{\fontsize{12}{16}\selectfont}
\subsectionfont{\fontsize{11}{14}\selectfont}
\subsubsectionfont{\fontsize{10.5}{12.5}\selectfont}

\newenvironment{packed_itemize}{
\begin{list}{\labelitemi}{\leftmargin=1em}
\setlength{\itemsep}{0pt}
\setlength{\parskip}{0pt}
\setlength{\parsep}{0pt}
\setlength{\headsep}{0pt}
\setlength{\topskip}{0pt}
\setlength{\topmargin}{0pt}
\setlength{\topsep}{0pt}
\setlength{\partopsep}{0pt}
\setlength{\itemindent}{0pt}%
\setlength{\leftmargin}{0pt}%
}{\end{list}}

\newenvironment{packed_enumerate}{
\begin{enumerate}
 \setlength{\itemsep}{2.5pt}
 \setlength{\parskip}{0pt}
 \setlength{\parsep}{0pt}
 \setlength{\headsep}{0pt}
 \setlength{\topskip}{0pt}
 \setlength{\topmargin}{0pt}
 \setlength{\topsep}{0pt}
 \setlength{\partopsep}{0pt}
}{\end{enumerate}}

\usepackage{hyperref}
\usepackage[T1]{fontenc}
\usepackage[nobiblatex]{xurl}
\usepackage{etoolbox}
  
\begin{document}

\title{Testing Access-Control Configuration Changes for Web Applications}

\author{
{\rm Chengcheng Xiang}\\
University of California, San Diego
\and
{\rm Li Zhong}\\
University of California, San Diego
\and
{\rm Eric Mugnier}\\
University of California, San Diego
\and
{\rm Nathaniel Nguyen}\\
University of California, San Diego
\and
{\rm Yuanyuan Zhou}\\
University of California, San Diego
\and
{\rm Tianyin Xu}\\
University of Illinois Urbana-Champaign
}

\maketitle

\begin{abstract}
Access-control misconfigurations are among the main causes of today's data breaches in web applications. 
However, few techniques are available to support \textit{automatic} and \textit{systematic} testing for access-control changes and detecting risky changes to prevent severe consequences.
As a result, those critical security configurations often lack testing, or are tested manually in an \textit{ad hoc} way.

This paper advocates that tests should be made available for users to test access-control configuration changes.
The key challenges are such tests need to be run with production environments (to reason end-to-end behavior) and need to be performance-efficient.
We present a new approach to create such tests, as a \textit{mini test environment} incorporating production program and data, called \acltest{}s.
\acltestcap{}s report the impacts of access-control changes---the requests, which were denied, would be allowed after a change, and vice versa. 
Users can validate if the changed requests are intended or not and identify potential security vulnerabilities.

We evaluate \acltest{}s with 193 public configurations of widely-used web applications on Dockerhub.
\acltestcap{}s detect 168 new vulnerabilities from 72 configuration images. 
We report them to the image maintainers; 54 of them have been confirmed and 44 have been fixed.
We also conduct in-depth experiments with five real-world deployed systems, including Wikipedia and a commercial company's web proxy.
Our results show that \acltest{}s effectively and efficiently detect all the change impacts.
\end{abstract}
\section{Introduction}
\label{sec:intro}

Access-control misconfigurations have been the root causes of many recent security incidents.
For example, in 2022, a misconfigured government website leaked confidential gun permit data~\cite{gunPermit}. 
The access control was only configured for the Tableau workbook, but was not configured for the sheets to prevent public access to the underlying dataset. 
In 2019, misconfigurations caused data leaks of election campaign websites~\cite{trumpCampaign};
in the same year, a misconfigured website of Inmediata Health Group exposed 1.5M patient data, causing disastrous privacy leaks~\cite{immediateHealth}.

One main challenge of access-control configuration management is to accommodate constant configuration changes.
Configuring access control is not a one-time effort but a continuous process to address the dynamics of systems and data policies.
In modern web applications, system components and features are continuously deployed and upgraded~\cite{duvall2007continuous, humble2010continuous, shahin2017continuous},
while data policies are rapidly created and updated~\cite{xu:chi:17,das2010baaz,sinclair:10}.

In our experience of managing access control of different web applications across companies (including some of the largest web applications in terms of end users), today's access-control configuration management still undergoes manual, often ad hoc practices.
Typically, when a legitimate access (to data or services) is denied, system administrators or DevOps engineers will need to manually reconfigure access control.
However, if access-control misconfigurations are introduced especially in the form of over-permission, the misconfigurations often remain latent for a long time, until being noticed by permission reviews, or detected by the red teams, or even worse revealed by true security incidents.

\begin{figure}[t!]
    \centering
    \includegraphics[width=0.46\textwidth]{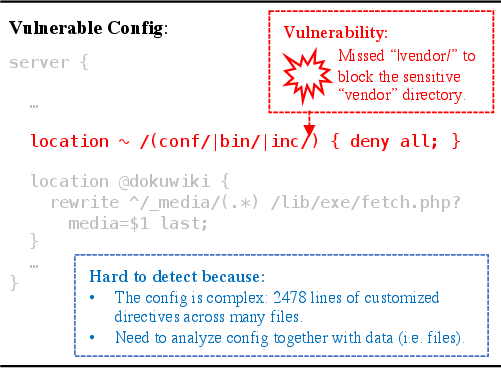}
    \caption{
    \textbf{A new access-control vulnerability we detected in a widely used Docker image (10M+ downloads)~\cite{linuxserver}. The vulnerability has been confirmed and fixed~\cite{linuxserver_report}. } 
    }
    \label{fig:intro-example}
\end{figure}

It is not surprising that existing human-based practices fall short in front of 
    the challenges of access-control configuration management.
The complexity of access-control configurations makes it hard for administrators or DevOps engineers to reason about the {\it impact} of access-control changes. Figure~\ref{fig:intro-example} shows a configuration snippet of the Dokuwiki web application---it is nontrivial to detect the vulnerability in the configuration files with 2,478 lines of complex directives.
Second, as reported repeatedly in prior studies~\cite{west:08,lampson:09}, to err is human and misconfigurations are inevitable, especially considering that access-control configurations are often done with high time pressure to unblock important accesses~\cite{west:08,lampson:09,xu:chi:17}.

We argue that a key missing piece of today's access-control configuration management is automated and continuous testing to evaluate the {\it impacts} of access-control changes and to flag problematic changes. 
In other words, changes of access-control configurations should be rigorously tested and evaluated, in the same way as source-code changes~\cite{adrion1982validation, huang1975approach} and other configuration changes for feature flags and performance tuning~\cite{tang2015holistic,sun2020testing,wang2023test,cheng2021test}. 
Unfortunately, we are unaware of
a practical, ready-to-deploy approach to effectively testing configuration changes to access control of modern web applications.

In fact, testing and verification were considered as promising directions to address access-control configurations a decade ago with specialized testing and verification techniques being developed~\cite{fisler2005verification, martin2007automated, martin2007fault, xu2018automated, xu2020automated}. However, the aforementioned techniques all assumed a generic, high-level configuration language, XACML~\cite{lorch2003first}, i.e., the target access-control configuration is encoded in XACML. Unfortunately, in practice, few web application component adopts XACML. Instead, access-control configurations today are often encoded in {\it system-specific} languages.
For example, NGINX does not use XACML but uses a specific directive language (Figure~\ref{fig:intro-example}).  
The semantics of NGINX's configurations are different from other web servers like HTTPD, Lighttpd, and Tomcat, and are also drastically different from other web system components such as front-end proxies and databases.
The prevalent uses of system-specific access-control configuration languages makes it hard to apply access-control testing and verification techniques for XACML, and it would take significant overhead to translate system-specific languages into XACML.

Note that existing configuration testing techniques~\cite{sun2020testing,wang2023test,cheng2021test} are fundamentally limited and are ineffective for access-control configurations.
First, existing techniques focus on unit and integration test, and cannot reason about end-to-end access behavior. 
More fundamentally, existing configuration testing are disconnected with application data in the deployment environment.
However, access-control configuration cannot be reasoned without the context of application data. 
In Figure~\ref{fig:intro-example}, the configuration is vulnerable because the \texttt{vendor} directory contains sensitive data with open access.
A test cannot detect the vulnerability without knowledge of the \texttt{vendor} directory on the file system.
It would be prohibitively expensive to address this issue by replicating production data in the test environment~\cite{testinproduction, testinproduction2}.

\vspace{3pt}
\noindent
{\bf Contributions.} In this paper, we present a general and practical approach of testing access-control configuration changes for web applications.
Our goal is to enable systematic evaluation of the impact of target configuration changes as a foundation to detect access-control vulnerabilities.

The key idea is to create a {\it mini test environment} named {\it \acltest{}} to efficiently and effectively evaluate the impact of every target change. The impact is
    represented by the differences of access-control decisions for user requests before and after the target change, i.e., what requested that were denied are now allowed by the configuration change, and vice versa.

Our key insight is that \acltest{} can be directly generated 
    from the target web applications
    and their deployment environments,
    without manually implementing new test code
    or replicating data and system dependencies,
    and the created ACtest is readily deployable and safe to run.

\acltest{} is an {\it ephemeral} test environment that are ``forked'' from the deployment environment with high fidelity (including the web application, the application data, and other dependencies).  
It allows safe data access (i.e. read but not write) from the test web application to the production data.
The safe data sharing are enabled at file system, database and network layer through techniques of Copy-on-Write, write protection and virtual subnet.

\acltest{} is efficient. It minimizes the original web application into a slim, fast-running application to speedup the test execution, using a novel program trimming technique.
The minimization is important, because with the original web application, the testing would take too long to run against comprehensive user requests (e.g., due to excessive system call and I/O operations commonly in web applications).
We found the minimization can be done systematically, because the expensive operations are typically only executed {\it after} access-control checks---skipping them does not affect testing results for access-control configurations.

\acltest{} measures the impact of the target access-control configuration change based on generating comprehensive user requests.
It supports (1) replaying the historical user requests recorded in access logs and calculating the differences of access-control results before and after the change, (2) synthesizing user requests that cover the Cartesian product of existing subjects (e.g., a user), objects (e.g., a file), and actions (e.g., a GET or SET API).

\vspace{3pt}
\noindent
{\bf Key results.} 
We evaluate \acltest{}s with the public Docker images of five widely-used web applications. 
By testing the changes made in the images version history,
\acltest{}s detect 168 new vulnerabilities from 72 Docker images\protect\footnotemark.
We have reported them to the image maintainers. 
So far 54 of them have been confirmed and 44 of them have been fixed by the maintainers.
We also evaluate \acltest{}s with five deployed web applications, including Wikipedia and the web proxy of a commercial company with millions of users. 
The results show (1) by synthesizing user requests, \acltest{}s detects all the impacts of injected changes; (2) by replaying historical user requests, \acltest{}s detects up to 80\% of the impacts.
The minimization reduces up to 98.61\% testing time.

\footnotetext{The full list of our detected vulnerabilities can be accessed anonymously through: \url{https://docs.google.com/spreadsheets/d/18_oaaoEawPJ-GNajqEL7t43tiKQ2iizZ9J6URCZMWc0}}

\vspace{3pt}
\noindent
{\bf Summary.} 
This paper makes three main contributions:
\begin{packed_itemize}
    \item We present ACtest as a new approach for testing access-control configuration changes and evaluating their impacts using mini, ephemeral test environment.
    \item We develop techniques to create ACtest from target web applications and their deployment environment, including safe data access and performance optimization.
    \item We show that ACtest can help effectively detect access-control misconfigurations by applying it to Docker images; ACtest has detected
    168 new vulnerabilities (54 confirmed and 44 fixed).
\end{packed_itemize}
\section{Motivating Examples}
\label{sec:motivation}

We present how change impacts detected by \acltest{}s can help sysadmins validate their access-control configuration changes and identify security vulnerabilities.
We use two real-world access-control misconfigurations \acltest{}s detected from Docker images as examples.

\subsection{Dangerous web interfaces}
\acltestcap{}s can detect unexpected interfaces and resources introduced by configuration changes.
This helps sysadmins validate a change when they install third-party plugins or extensions, a common practice for web applications~\cite{wordpresspluginpage, mediawikiextensionpage,drupalmodulepage}.
As exemplified by both real-world incidents~\cite{wordpressplugin} and our evaluation on Docker images,
installing extensions could incur security risks.

\begin{figure}[t!]
    \centering
    \includegraphics[width=0.48\textwidth]{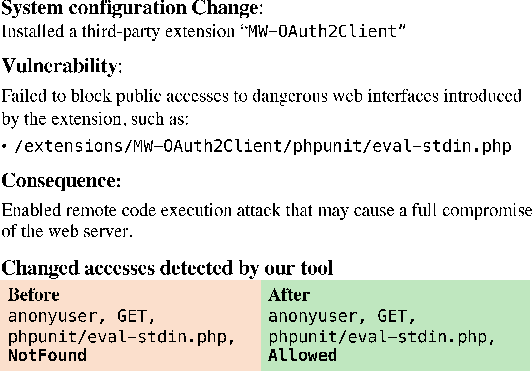}
    \caption{
    \textbf{A new vulnerability detected by \acltest{}s in a third-party Docker image~\cite{servarr} of MediaWiki~\cite{mediawiki}.} 
    This vulnerability has been confirmed and fixed by the image maintainer\protect\footnotemark. 
    \acltestcap{}s detect that anonymous users cannot access \texttt{phpunit/eval-stdin.php} by default, but can access it after the change.
    }
    \label{fig:example1}
\end{figure}

Figure~\ref{fig:example1} shows a new vulnerability detected by \acltest{}s in a Docker image of MediaWiki~\cite{mediawiki}, a popular open-source wiki system.
In this example, the sysadmin installed a third-party PHP extension ``\texttt{MW-OAuth2Client}'' that introduced several dangerous PHP files, including ``\texttt{eval-stdin.php}''.
These files exposed web interfaces that should only be used for testing environments and can be exploited for remote code execution attacks in production.
However, sysadmins were not aware of these files and failed to apply any access-control rules to limit the access, making the dangerous interfaces in the files publicly accessible.
\acltestcap{}s effectively detect that these files were not accessible before the installation, but become accessible after. It warns sysadmins so they can double check if the exposed interfaces in the files are safe and make appropriate configurations to prevent requests to the dangerous interfaces before they are exploited.

\footnotetext{Our report: \url{https://github.com/Servarr/mediawiki-docker/issues/1}; reproducing: \url{https://github.com/conf-test/acl-test/blob/main/reproduce-example.md}}

\subsection{Openly-accessible database dump}
\acltestcap{}s can also detect access-control misconfigurations that lead to unexpected impacts.
Oftentimes, access-control configurations are complex and error-prone.
It is necessary to test the actual impacts in terms of system behavior for sysadmins to validate if the behavior change meets their intention.

\begin{figure}[t!]
    \centering
    \includegraphics[width=0.48\textwidth]{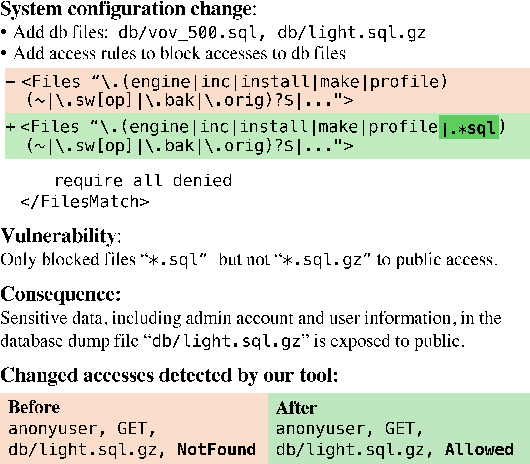}
    \caption{
    \textbf{A new vulnerability detected by \acltest{}s in a third-party Docker image\cite{vulfocus} of Drupal~\cite{drupal}.}
    \acltest{}s detect that anonymous users cannot access the db dump file \texttt{db/light.sql.gz} before but can access it after the change\protect\footnotemark.
    }
    \label{fig:example2}
\end{figure}

Figure~\ref{fig:example2} shows a new vulnerability detected by \acltest{}s in a Docker image of Drupal (a widely-used CMS system).
In a change, the sysadmin adds a few database files, such as 
``\texttt{vov\_500.sql}'' and ``\texttt{db/light.sql.gz}''.
She adds a customized access-control rule to prevent these files from being accessed publicly.
The rule is in the form of a regular expression ``\texttt{*sql}''.
However, the added rule only blocks files with ``\texttt{.sql}'' but not with ``\texttt{.sql.gz}''.
This makes the file ``\texttt{db/light.sql.gz}'' open to public access.
Note that this file is a database dump, including both users data and admins account info.
\acltestcap{}s effectively detect only the ``\texttt{.sql}'' files are blocked but the ``\texttt{.sql.gz}'' files are still accessible.
It can warn sysadmins so that they can further change their configurations to block requests to the ``\texttt{.sql.gz}'' files.

\footnotetext{We reported this through email.}

\section{\acltestcap{} Overview}
\label{sec:overview}
\acltest{} is a mini test environment for helping sysadmins analyze the {\it impact} of access-control changes in terms of end-to-end system behavior.
\acltest{} analyzes access-control changes by running test with production programs, configurations and data. 
\acltestcap{} outputs the impact of configuration changes, in the form of access-control behavior comparison.
Based on \acltest{}'s output, sysadmins can examine whether the configuration change results are intended access-control behavior. \acltestcap{} can help sysadmins to identify incorrect or unintended access-control that either leads to security vulnerabilities such as data breaches (grant more access than what is intended) or accessibility issues (grant insufficient access for the desired functionality).

\subsection{\acltestcap{} Definitions}
\label{sec:def}
For a target system, an \acltest{} is a system-level test environment, denoted as $t(P, C, D)$. It consists of: 
1) the system program $P$,
1) the production configuration $C$, 
2) the production data $D$. 
Before a configuration change $\Delta C$ is rolled out to production, \acltest{} runs $P$ twice, with $C$ and $D$ as well as with $C+\Delta C$ and $D$ to evaluate the end-to-end behavior change before and after $\Delta C$ is rolled out.
Note that $C$ and $D$ are the entire set of system configurations and data, not limited to access-control related ones.
As discussed before, this is needed to achieve high-fidelity of the testing results. 
Since production data can have large volumes, the \acltest{} does not replicate them for testing but access production data in a safe way (\$\ref{sec:transform-safe}).

After testing a configuration change $\Delta C$, the \acltest{} outputs the change impact as a set of requests that have different results before and after applying $\Delta C$---namely, requests that were denied by $C$, but become allowed after applying $C+\Delta C$, or vice versa.
A request is a tuple $\langle s,o,a,r \rangle$, which represents a subject $s$ (e.g., a user) that performs an action $a$ (e.g., an HTTP GET/PUT request) to an object $o$ (e.g., a file) and gets a result $r$ (either ALLOW or DENY).
Therefore, the change impact is a set of tuples $\langle s,o,a,r, r' \rangle$, where $r$ is the access results with $C$, $r'$ is the result with $\Delta C$ and $r\neq r'$.
Our definition of access-control change impact is consistent with prior work~\cite{xiang2019towards}.

To obtain the change impact, \acltest{}s need to generate test inputs to feed into $P$ to find out the behavior change.
\acltestcap{}s provide two ways to generate test inputs, in the form of $\langle s, o, a \rangle$.
First, \acltest{}s allow sysadmins to provide access logs which record the historical requests.
The access logs help \acltest{}s cover the common requests.
Second, \acltest{}s also allow sysadmins to specify the subjects, objects, and actions, which will be used to generate all possible combinations to achieve a complete coverage.

\subsection{Usage and Deployment Model}

\acltest{}s consists of production program, configuration and data.
To set up \acltest{}s , sysadmins only need to specify a few inputs: 
1) the location of the production program $P$, configuration $C$ and data $D$;
(2) the target configuration change $\Delta C$;
3) a set of access logs, and/or a specification of target subjects, actions and objects (e.g., the root directory of all HTML files to test);
and 4) optionally, the limit of CPU and memory that \acltest{}s are expected to take.
Note that 1), 2) and 4) have already been commonly done in today's deployment practices with container images (e.g., those on DockerHub~\cite{dockerhub}). Thus the true additional work to run \acltest{}s is only 3).

Multiple \acltest{} instances may need to be deployed to test different programs.
Multiple \acltest{} instances can be placed in a single virtualized subnet so they can communicate with each other but do not affect production processes (cf. \S\ref{sec:safenetwork}). 
Multiple \acltest{} instances can also be configured with some of them running trimmed version while others not. This is necessary when testing one instance requires the full execution of another instance. For example, to test access control implemented in PHP code, only the PHP code can be trimmed but not the Httpd server; otherwise, the PHP code won't be executed.

Sysadmins can choose to run \acltestcap{}s when they know a change is going to be made, or run \acltestcap{}s periodically and get notified with behavior changes. 

Compared with setting up a mirror environment for testing, \acltest{}s save sysadmins' effort of maintaining test configuration $C$ and data $D$, as well as test running time. 
\S\ref{sec:transform-safe} will present how \acltest{}s safely access production configuration and data without causing side effect to them.
\S\ref{sec:optimization} will present how target programs are trimmed to accelerate the execution of \acltest{}s.

\section{\acltestcap{} Generation}
\label{sec:design}
We present a general approach, \acltestgen{}, to help developers create \acltest{}s from and for their web applications.
To achieve this, \acltestgen{} addresses three challenges:
\begin{packed_itemize}
    \item How to create \textit{safe} \acltest{}s from original applications? Since \acltest{}s will run in production, 
    they cannot incur side effects to production environments.
    \item How to make \acltest{}s \textit{performance-efficient}? \acltestcap{}s may need to test a large number of requests (the combination space of subjects, objects and actions can be large). 
    \item How to \textit{minimize users' effort} including effort on providing test inputs and on validating test results?
\end{packed_itemize}

To make \acltest{}s safe to run in production, \acltestgen{} systematically enable \acltest{}s' safe access to different types of data.
For file access,  \acltestgen{} sets up \acltest{}s in a copy-on-write file system.
For database access, \acltestgen{} grants read-only permission for \acltest{}.
For network access, \acltestgen{} isolates \acltest{} in a virtual subnet.

To improve test performance, \acltestgen{} uses a novel trimming technique to modify an original program to make it run faster. 
The key idea is that access-control checks (ACCs) are usually run before the main task execution in a program run. Therefore, the main task execution can be skipped without impacting the results of ACCs.

To minimize effort on providing test inputs, \acltest{}s are generated with two methods for users to specify how to generate test requests: using access logs and letting users specify subjects, objects and actions.
To minimize effort on validating test results, \acltest{}s aggregate impacted requests by attributes of subjects, objects and actions.

\subsection{Creating \textit{Safe} \acltestcap{}s}
\label{sec:transform-safe}

To evaluate the access-control behavior of a system with high fidelity, the best way is to test the access-control configurations with the production application and environments.
However, it is unsafe to test the application directly with production environments, because it can cause side effects, like writing to production data.

\begin{figure}[t!]
    \centering
    \includegraphics[width=0.4\textwidth]{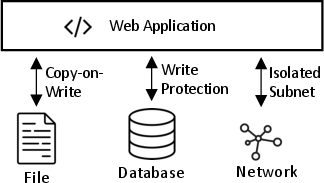}
    \caption{
    \textbf{Safe data access in \acltest{}.}
    }
    \label{fig:safety}
\end{figure}

To eliminate the side effects, \acltestgen{} provides systematic mechanisms to handle different types of data accesses, as shown in Figure~\ref{fig:safety}. 
A web application accesses data mainly through three ways: file system, database and network.
Reading from these production data is necessary for the web application to produce real access-control decisions; 
however, writing to the production data is not acceptable for testing.

\subsubsection{Safe File Access}
\label{sec:fileaccess}
Web applications need to read files for loading configurations (e.g. `.htaccess`) or web pages (e.g. HTML or PHP files), and needs to write files for editing or uploading files. 
\acltestgen{} adopts file system Copy-on-Write (CoW) techniques~\cite{overlayfs, rodeh2013btrfs} to allow reading from production files while avoid writing to production files.
With CoW, the production files are directly shared to \acltest{}s for all read requests, while a copy is made when \acltest{}s try to modify any production configurations and data. 
The modification is only applied to the copied version, which is invisible to production systems, so it does not affect the execution of production systems.

In our implementation, \acltestgen{} sets up an \acltest{} with OverlayFS~\cite{overlayfs}, a union mount filesystem supported by Linux kernel mainline and Docker. 
The CoW mechanism may slow down \acltest{}s' execution as writes to files may need to copy them first. 
Luckily, read operations dominate most server workloads~\cite{nishtala2013scaling, armstrong2013linkbench, cao2020characterizing, harter2014analysis, terrace2009object, fetterly2011tidyfs}.
In addition, we will present in \S\ref{sec:optimization} our method for removing I/O related code that is irrelevant to access-control from the original program.

\subsubsection{Safe Database Access}
Web applications needs to access databases for querying or updating configurations (e.g. user permission tables) as well as data (e.g. web page content tables).
To let web applications generate real access-control results, it is necessary to have the database queries return production data to applications.
\acltestgen{} provides a write-protection mechanism to allow query production databases while block update production databases.

There are multiple ways to implement write-protection in modern database management systems (e.g. MySQL, PostgreSQL).
First, we can take advantages of databases' own permission systems.
Modern databases usually provides read/write permission roles to different clients. 
To block write while allow read, 
we can grant web applications a read-only role. 
Second, we can also use file system's CoW for database write-protection.
At the storage level, database tables are also stored as data files in file systems.
We can mount the production data files in a CoW file systems, so that the database will read from production data but will write to a shadow copy.

In our implementation, \acltestgen{} take the file system CoW approach for database write-protection.
\acltestgen{} sets up a test database instance which has it db data files mounted from the production db data files with CoW enabled.
A dedicated database instance for test is necessary, as we do not want the test database's in-memory states interfere with the production database's in-memory states.

\subsubsection{Safe Network Access}
\label{sec:safenetwork}
Web applications may need to communicate with other services (e.g. database instances, in-memory cache services) to obtain the data it needs.
However, it is problematic to let an \acltest{} instance talk with production services as it may cause an expected state change of the production services.

To address the issue, \acltestgen{} locates \acltest{} instances in a virtualized subnet. 
Within a virtualized subnet, an \acltest{} instance can only talk with other instances in the same subnet. 
To create a complete mini test environment for web applications,
\acltestgen{} usually need to set up two \acltest{} instances/services, including the web server and database, in a same subnet.
Note that not all services in production is required for access-control testing. 
For instance, a cache service can be removed from testing environment, as it only affect performance but not access-control results.

\subsection{Making \acltestcap{}s Performance-efficient}
\label{sec:optimization}

Running a whole original program for testing can take a long time and consume a lot of resources.
We developed a novel trimming technique to modify the original program to accelerate test execution.

\begin{figure}[t!]
    \centering
    \includegraphics[width=0.46\textwidth]{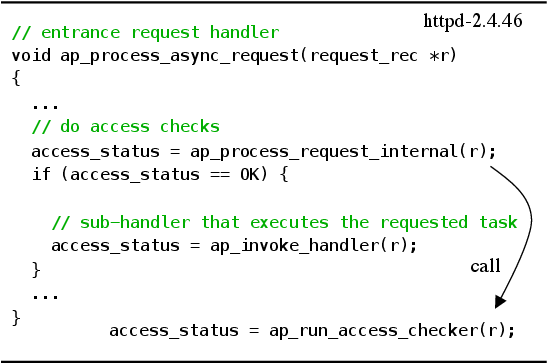}
    \caption{
    \textbf{Code pattern of access-control checks (ACCs).}
    Each request is processed by an entry handler, which first does ACCs and then calls a sub-handler to execute the requested task. 
    Strawman trimming removes the sub-handler call completely. 
    }
    \label{fig:accpattern}
\end{figure}

Our trimming technique is based on two code patterns we observed about the access-control checks (ACCs) in server programs and web applications:
\begin{packed_itemize}
\item ACCs are executed in the early phase of server programs' request handlers. Server programs call request handlers to process incoming requests, and request handlers run ACCs before executing the actual requested task that takes up the most execution time.
\item The results of ACCs are used as a condition to decide if the requested task will be executed. Only when the results are ``allow'', the requested task will be run.
\end{packed_itemize}
We validate these patterns in five widely used server programs and web applications: Httpd, Nginx, Wordpress, Mediawiki and Dokuwiki.
The patterns hold with all of them.
An example of the code pattern is shown in Figure~\ref{fig:accpattern}. 
The overall handler called for each request is denoted as an entry request handler and the actual requested task is executed in a sub-handler.

Based on the observed code patterns, we develop trimming techniques to skip the execution of the actual requested task, which is time-consuming.
We start with a strawman approach to remove sub-handlers in whole, which unfortunately also remove some ACCs. 
Then we design an advanced trimming algorithm to adopt hybrid program analysis for fine-grained trimming.

\subsubsection{Strawman trimming}
\label{sec:basicidea}

Ideally, all the ACCs have been done before the sub-handlers that execute the actual requested tasks.
In this case, the access-control results have been decided before the sub-handlers.
Since our testing goal is only about the access-control results, our strawman approach removes the sub-handlers in total.

However, this strawman approach can lead to wrong access-control results. The reason is that the sub-handlers may also perform additional task-specific ACCs.
Our study shows that there are two types of task-specific checks: 1) file permissions---programs typically check file permissions only when they try to open files;
2) task-specific configurations---depending on what kind of tasks to execute, sub-handlers may have further ACCs differing from the global checks in the entry request handlers. 
For example, different HTTP requests in Apache need to check for different permissions like for proxy redirection, cgi script execution and directory listing.

We experimented with this strawman trimming. 
As we will present in \S\ref{sec:exp_inject}, 
for four systems, \acltest{}s generated with strawman trimming lead to false access-control results: 
three of them can detect 46.4\% - 52.7\% of the change impacts, and the remaining one can detect no change impact.

\subsubsection{Advanced Trimming}
To prevent ACCs from being trimmed, it is necessary to delve into each sub-handler to perform a fine-grained trimming. 
Such a fine-grained trimming needs to address two challenges:
1) How to find the final ACCs so that the trimming will not remove any ACCs before them?
2) How to safely trim the code after the final ACCs so that the trimmed program can be compiled and run correctly?

\vspace{8pt}
\noindent \textbf{Finding final ACCs with static analysis.}
Our first attempt is to build a static analysis tool to find the final ACCs that are executed in a program's execution paths.
The idea is to build an inter-procedural control-flow graph (CFG) for a whole program, traverse all the paths from the entry node (the main function), and identify the final ACCs in the paths.
This approach turns out not to work because of three reasons.
First, the number of paths in a whole-program CFG explodes, and so the analysis cannot scale to real-world programs (e.g. Httpd, Nginx, Wordpress and Mediawiki in our study).
Second, the inter-procedural property makes the analysis fragile to function pointers.
Function pointers set at run time make it hard to statically decide functions' accurate execution order. And without knowing the accurate execution order, it is not possible to decide which ACCs are the final ones.
Third, the analysis needs the knowledge of ACCs, which can be various and can hardly be specified manually.

\vspace{8pt}
\noindent \textbf{Finding final ACCs with hybrid analysis.}
To address the challenges, we incorporate dynamic analysis with static analysis to form a hybrid approach.
First, instead of building a static CFG for a program, our approach runs the program with various configurations and builds dynamic CFGs for it.
Second, since the CFGs are built dynamically, the function pointers are decided and so the execution order is known.
Third, we designed a novel CFG diff algorithm to compare the CFGs of the allowed runs (where requests are allowed) and denied runs (where requests are denied) to identify ACCs automatically.

Our hybrid analysis takes three steps:
1) Dynamic analysis---it instruments a program, runs the program multiple times with various configurations as well as requests, and builds dynamic CFGs from the runs;
2) CFG-diff---it compares the dynamic CFGs to identify the final ACCs in them;
3) Static analysis---starting with the dynamic CFGs and the identified final ACCs, it performs a static analysis to find the final ACCs that have not been included in the dynamic CFGs. We detail each steps as following:

\textit{1) Dynamic analysis:} Our dynamic analysis takes tuples of $\langle r, C, C'\rangle$ as inputs to run a program, where $r$ is a test request, $C$ is a configuration that allows the request and $C'$ is a small mutation to $C$ that denies the request.
Our analysis runs the program with $C$ and $C'$ separately to handle $r$ and build a pair of dynamic CFGs for step 2.
The number of the inputs decides how many execution paths can be trimmed; therefore, a good set of $\langle r, C, C'\rangle$ inputs is expected to exercise the access-control of a program's main functionalities.

We explored multiple ways to obtain the input for dynamic analysis. At first, we expected that such inputs should be the subset of programs' functional tests, and we tried to find existing functional tests for our analysis.
However, we only found very few and incomplete functional tests for the programs we studied.
As a result, we turned to software manuals to create $\langle r, C, C'\rangle$s by ourselves.
Fortunately, most manuals~\cite{nginxmanual, apachemanual, postgresmanual, vsftpdmanual} give clear guidance on how to set up and test the main functionalities. 
Under the guidance of the manuals, we successfully create test inputs for our experiments (the scripts for creating inputs in our experiments can be found online~\cite{nginxtriminput,apachetriminput,vsftpdtriminput,postgrestriminput}). 
Our experience shows that it is not hard for even non-developers of a program (like us) to create the inputs to cover the main functionalities.
Note that an incomplete set of inputs is acceptable as this affects only the effectiveness of trimming (trim fewer execution paths) but not the correctness (trim ACCs by mistake).
Static analysis in step 3 is designed to prevent an ACC from being trimmed even if it is not exercised by the inputs.

\begin{figure*}[t!]
    \centering
    \includegraphics[width=0.98\textwidth]{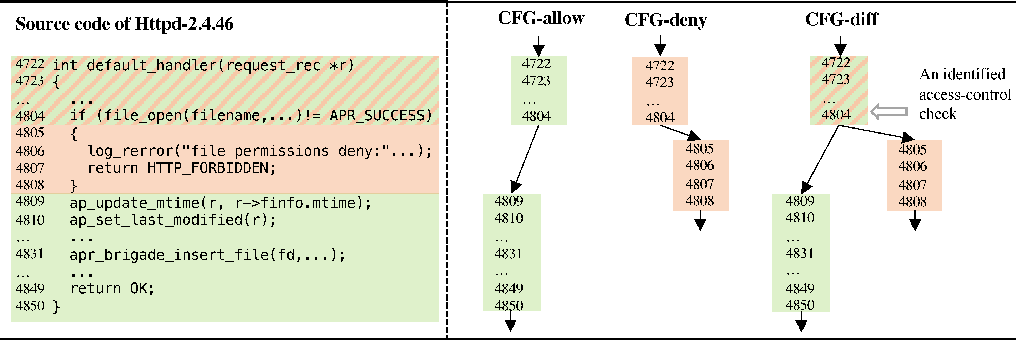}
    \caption{
    \textbf{An example of using CFG-diff to identify ACCs.}
    \textbf{CFG-allow}: dynamic CFG for a run allows an request; 
    \textbf{CFG-deny}: dynamic CFG for a run denies an request;
    \textbf{CFG-diff}: a merged CFG with node colored differently.
    The examples are simplified.
    Complete CFG-diffs generated in our experiments can be found online~\cite{cfg-diff}.
    }
    \label{fig:finetrim}
\end{figure*}

\textit{2) CFG-diff:} We propose a novel CFG-diff algorithm to identify ACCs. The algorithm is based on an observation that each pair of allowed and denied CFGs from the previous step should diverge mainly on the final ACC.
An example is shown in Figure~\ref{fig:finetrim}, where the \texttt{file\_open} check diverges the CFGs. 
To identify the diverging node, CFG-diff merges CFG-allow and CFG-deny into one CFG, colors the nodes differently, 
and finds the nodes with mixed color (``green-red-mixed'') but has differently-colored children (``green'' and ``red''), 
as shown in Algorithm~\ref{alg:cfg-diff} in appendix.

\textit{3) Static analysis:} CFG-diff finds a final ACC in each pair of allowed and denied dynamic CFG. 
However, the final ACCs found are not complete as the dynamic CFGs may not contain all the final ACCs.
And the final ACCs found may also not be the true final ones as a final ACC in one pair of CFGs may appear before another ACC in another pair of CFGs.
To address this, we further incorporate static analysis, including a forward analysis and a backward analysis, as shown in Algorithm~\ref{alg:forward_backward} in appendix.
For each final ACC found by CFG-diff, the forward analysis \textit{adds} the ACCs performed after it to the candidate final ACCs, while the backward analysis \textit{deletes} the ACCs performed before it from the candidate final ACCs.
The forward and backward analysis also expand the dynamic CFGs statically so that the new ACCs that are not executed in previous dynamic analysis can also be found.
To identify the new ACCs, our analysis adopts the knowledge of check function names and the return types (e.g. the \texttt{ACCESS\_DENY} enum) learned from the identified ACCs from step 2.

\vspace{8pt}
\noindent \textbf{Trimming programs safely.} To ensure the trimmed program can be correctly compiled and run, we designed a trimming method that makes minimal changes to the original program.
The idea is to modify the program so it always executes the access deny branches after the final ACCs no matter whether the checks return ``allow'' or ``deny''.
As access deny branches usually finalize a request handling quickly, 
this saves much execution time compared with executing the allowed branches.

\begin{figure}[t!]
    \centering
    \includegraphics[width=0.45\textwidth]{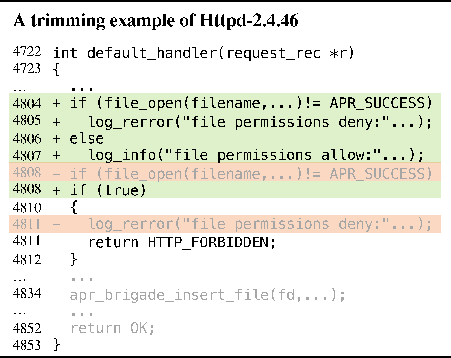}
    \caption{
    \textbf{Safe trimming for an ACC in Httpd-2.4.46}:
    1. The access deny branch condition is set to be always true;
    2. The original ACC is modified only to control logging.
    }
    \label{fig:finetrim2}
\end{figure}

Figure~\ref{fig:finetrim2} presents an example of how our trimming method works. 
First, it modifies the access-deny branch condition to constant \texttt{true}.
Second, it uses the original ACCs only for logging the access-control results.
Last, it advances the access-control logging statements after the original checks to the new checks, which makes sure the correct access-control results are recorded.

\subsection{Minimizing Usage Effort}
To minimize the usage effort of \acltest{}s, we design three techniques to generate test inputs with little human specification, to simulate test requests automatically, as well as to present test results effectively and perform automatic triage.

\subsubsection{Generating Test Inputs}
\acltest{}s provides two options for sysadmins to specify what requests to generate.
The first one accepts a set of historical access logs along with annotations of the subject, object and action fields.
\acltest{}s parse access logs into $\langle s, o, a\rangle$ tuples and uses them as the test requests.
The second option accepts the sources of subjects, objects and actions in a current system. 
Such sources include database tables that store users ($S$), directories that store all file paths ($O$) and a list of all actions ($A$). 
\acltest{}s then traverse the Cartesian product of the $S \times O \times A$ to synthesize the test requests.
Depending on sysadmins' specification, the synthesized requests can be either the one related to a change or be all possible requests.

\subsubsection{Simulating Requests}
\acltestcap{}s includes two simulating techniques.
The first one is simulating requests from various IP addresses. 
This is commonly needed as many access-controls are based on IP addresses, like blocking certain ones.
We implement this with the AnyIP~\cite{anyip} feature provided by Linux Kernel.
This feature enables our testing container to bind the whole IPv4 or IPv6 address space to its loopback device.
In this case, a request sent to any third-party IP address will be routed back to itself and will be recognized as from the third-party IP address.
This enables the server program to do access-control with the correct source IP.
The second one is concurrent requests. 
Sending concurrent requests is essential for accelerating the testing, besides the trimming method we proposed.
We implement a concurrent framework which uses multiple processes and multiple threads to sends requests.
We take into account the dependencies between requests from the same IP or the same user so that these requests will not be concurrently sent.

\subsubsection{Presenting Results And Performing Automatic Triage}
The number of impacted requests can be large even with a small configuration change.
Such examples include allowing access to a directory with many files.
The configuration change may only be a small change to the directory permission,
but the impacted requests include the ones to all the files under the directory.
Asking sysadmins to validate every access is time-consuming.

\acltestcap{}s present the impacted requests in an efficient way to assist sysadmins to validate them.
\acltestcap{}s aggregate impacted requests based on the common attributes of subject (user name and group), object (directory name and file suffix) and action (action type).
For example, if requests to all files under a directory are impacted, \acltest{}s only show the directory name to sysadmins.
\acltestcap{}s also show the file suffix impacted under the directory, which may raise sysadmins' attention to dig more.
The aggregation can be easily extended to other attribute types and makes the validation efficient.

\acltestcap{}s also perform automatic triage to classify the impacted requests into dangerous and less dangerous.
This allows sysadmins to only validate the dangerous ones if they have limited time.
Currently, \acltest{}s use a few conservative rules to identify dangerous change impacts: 1) the object name starts with a dot (i.e. hidden system files); 2) the object name ends with certain suffix (e.g. ".sql" for db dump); 3) the object name contain certain text (e.g. "phpunit" for unit test files); 4) the access method are dangerous (like "TRACE" for http).
Sysadmins can easily add or adjust the rules.
\subsection{Implementation}
We implement \acltestgen{} with existing container techniques.
First, a Dockerfile~\cite{docker-file} is generated to pack a target program into a container image.
Second, an overlay network~\cite{overlaynetwork} is set up for the image to make it run in an isolated subnet.
The name and IP address of the overlay network are generated as configurable parameters for sysadmins to flexibly place multiple \acltest{}s in a single subnet.
Third, OverlayFs~\cite{overlayfs} is configured to allow \acltest{}s access production data safely through Copy-on-Write.

We implement our trimming algorithm both for C/C++ and PHP programs. 
Our implementation takes the source code of a C/C++ or PHP program as input, and generates a trimmed version of the program for testing.
For C/C++ programs, we build a tool based on the LLVM framework~\cite{lattner2004llvm} to do dynamic and static analysis: 
1) We implement an LLVM transform pass~\cite{llvm-transform} to inject code to dynamically trace the execution of function calls and basic blocks;
2) We implement an LLVM analysis pass~\cite{llvm-analysis} to do intra-procedural static analysis.
For PHP program, we use Xdebug~\cite{xdebug} to do dynamic tracing and use PHP-Parser~\cite{PHP-Parser} to generate abstract syntax tree (AST) for static analysis.

\section{Evaluation}

\subsection{Methodology and Testing Environment}

We conduct three sets of experiments to measure \acltest{}s' effectiveness and efficiency:

\begin{packed_enumerate}
    \item Detecting real-world vulnerabilities: We use \acltest{}s to analyze the impacts of configuration changes in public Docker images and detect vulnerabilities.
    \item Controlled experiments: We replicate five real-world web systems, including Wikipedia and the web proxy of a commercial company. 
    We inject access-control configuration changes to them and use \acltest{}s to detect the change impacts.
    \item Performance: We run \acltest{}s with different trimming options to test the real systems as in the controlled experiments and compare their performance.
\end{packed_enumerate}

\subsection{Detecting Real-world Vulnerabilities}
\label{sec:detect_real_world_vul}

This experiment aims to evaluate \acltest{}s' effectiveness on detecting misconfiguration vulnerabilities.
We choose five widely-used web applications: Wordpress~\cite{wordpress}, Drupal~\cite{drupal}, Joomla~\cite{joomla}, Dokuwiki~\cite{dokuwiki} and Mediawiki~\cite{mediawiki} as the evaluation targets.
The images of each application were downloaded more than 10 million times on Dockerhub~\cite{dockerhub}.

For each system, we download the images updated recently and extract their configurations as well as data for testing. 
These configurations include Apache configurations, Nginx configurations and file permissions of web pages.
We use the official images' configurations as the initial version and the third-party images' configurations as the changed version (the official images are distributed by application vendors, while the third-party images are uploaded by application users who made some changes to the official ones).
\acltest{}s are generated to test the change impact of each changed version against the initial version. 
As historical logs are not available for these systems, \acltest{}s are set to synthesize requests regarding to all the combinations of users, actions and resources. 

The experiment focuses on the impact of risky changes---requests that are not allowed in the official images but are allowed in the third-party images. 
As not all the detected request changes are vulnerabilities, we manually validate them. 
We use a conservative standard to make the decision:
if the same type of allowed requests was identified as a vulnerability in \textit{other} systems or public vulnerability databases, we count the corresponding change as a vulnerability.
When multiple requests are impacted by the same change (e.g. multiple file requests are allowed because of a change to the parent directory), we only count it once.

\begin{table}[t!]
{
\centering
\setlength\tabcolsep{3pt}
\begin{tabular}{lccc}
\toprule
{\bf Systems}  & {\bf Images} & {\bf Vulnerable images } & {\bf Vulnerabilities}    \\
\midrule
Dokuwiki  & 57 & 19  & 67  \\
Mediawiki & 47 & 23  & 56  \\
Wordpress & 42 & 18  & 28  \\
Drupal    & 29 & 8  & 11  \\
Joomla    & 18 & 4   & 6  \\
\midrule
Total     & 193 & 72 & 168  \\
\bottomrule
\end{tabular}
\caption{
\textbf{ New vulnerabilities detected by \acltest{}s in public Docker images of popular web applications.} 
54 of the vulnerabilities has been confirmed~\cite{full_list_vul}. Many of the confirmed vulnerable images are widely-used: one image has more than 10 million downloads on DockerHub, six images have thousands of downloads each\protect\footnotemark{}. We were also able to find the online websites correspond to some of the vulnerable images\cite{BrAPI,nlpub.ru}. 
}
\label{tab:detectedmisconfigs}
}
\end{table}

\vspace{8pt}
\noindent \textbf{Overall Results.}
\acltestcap{}s reports \numofdetectedvul{} changes as dangerous vulnerabilities from \numofvulversions{} public Docker images~\cite{full_list_vul}, as shown in Table~\ref{tab:detectedmisconfigs}. 
We reported \numofreportedvul{} vulnerabilities to the maintainers of the images for which contact information is available. 
So far, 54 of the vulnerabilities have been confirmed and 44 of them have been fixed by the maintainers.
Examples of the detected vulnerabilities can be found in Figure~\ref{fig:example1} and \ref{fig:example2}.

\footnotetext{Example popular vulnerable 
images: \url{https://hub.docker.com/r/linuxserver/dokuwiki}, our report: \url{https://github.com/linuxserver/docker-dokuwiki/issues/33}
image: \url{https://hub.docker.com/r/touch4it/drupal-php-fpm-nginx}, our report: \url{https://github.com/touch4it/docker-php7/issues/18}; image: \url{https://hub.docker.com/r/mwaeckerlin/dokuwiki}, our report: \url{https://github.com/mwaeckerlin/dokuwiki/issues/1}; image: \url{https://hub.docker.com/r/vincowl/dokuwiki}, our report: \url{https://github.com/vincowl/dokuwiki-dockerfile/issues/1}}

In total, \acltest{}s detected 874 risky access changes from \numoftestedversions{} tested images.
Besides the 168 dangerous changes, \acltest{}s report the remaining 706 changes as less dangerous.
Those changes can also be potential vulnerabilities.
They expose system-specific resources and \acltest{}s cannot decide if they expose sensitive information or not. 
Therefore, sysadmins may choose to review them or search for sensitive keywords from them.

\begin{table}[t!]
{
\centering
\setlength\tabcolsep{4pt}
\begin{tabularx}{\linewidth}{>{\RaggedRight\arraybackslash\hsize=.43\hsize}X
>{\RaggedRight\arraybackslash\hsize=.9\hsize}X
>{\RaggedRight\arraybackslash\hsize=.24\hsize}X}
\toprule
{\bf Types of vul}  & {\bf Examples} & {\bf \# of vul}    \\
\midrule
Dangerous interface exposure  & Internal PHP interfaces for testing,  such as \texttt{phpunit/*} are exposed.  & 46  \\
\hline
Sensitive system settings exposure    & System settings like \texttt{composer.json} and \texttt{.htaccess} are exposed. & 40   \\
\hline
Sensitive metadata exposure   & Version control data like \texttt{.git/*} are exposed  &  16 \\ 
\hline
Dangerous access method enabled    & HTTP diagnostic method \texttt{TRACE} is enabled for public access. &   14 \\ 
\hline
Database files exposure  &  \texttt{*.sql} files are allowed to public download.  &  10 \\ 
\bottomrule
\end{tabularx}
\caption{
\textbf{Most common types of vulnerabilities detected.} }
\label{tab:vultypes}
}
\end{table}

\vspace{8pt}
\noindent \textbf{Vulnerability Types.}
Table~\ref{tab:vultypes} shows the common types of the detected vulnerabilities.
Many of the types are rated as high and medium severity~\cite{vul_composer, vul_git, vul_vendor, vul_htaccess}.
The most common type is the exposure of dangerous web interfaces.
Such examples include the exposure of PHP interfaces under directory ``\texttt{phpunit}'' and ``\texttt{vendor}'' to public access.
These interfaces may disclose sensitive data and allow arbitrary code execution~\cite{vul_phpunit, vul_vendor}.
These interfaces are located in third-party extensions and are introduced when sysadmins install the extensions.
As sysadmins have little knowledge of what a third-party extension may introduce, it is hard for them to be aware of the exposure of these interfaces.
However, with \acltest{}s reporting the impacts, sysadmins can inspect the impacts when they install an extension and can disable the exposures.

\vspace{8pt}
\noindent \textbf{False Positives.}
Besides the 54 vulnerabilities that were confirmed by the maintainers, eight reported vulnerabilities were marked as false positives by the maintainers.
We analyzed the false positives, and here are the reasons:
1) The image maintainers expect their users to further change the permissions to block the data exposure, as they have documented in their wizard. So they don’t consider the data exposure in their settings as vulnerabilities~\cite{linuxserver_fp};
2) The image maintainers think that the data exposure is not a problem because they are used in their intranet;
3) The image maintainers think the exposed data is already public, so is not a threat~\cite{EliasHolzmann_fp};
4) Bugs in our implementation identifies false changes. We have fixed the bugs~\cite{tkw1536_fp}.

\subsection{Detecting Change Impacts}
\label{sec:exp_inject}

This experiment aims to evaluate \acltest{}s' effectiveness on detecting the impacted requests of various access-control configuration changes in real-world deployed systems. 
We collect configurations and data from five real-world deployed web systems, including the Wikipedia website, a web proxy of a commercial company, a department course website, a research center website and a research group website.
We also collect access logs of them for \acltest{}s' test request generation.
For Wikipedia, because the whole website is too big for our testing environment, we use a part of it, called cowiki~\cite{cowiki}, as our testing target.

\begin{table}[t!]
{
\setlength\tabcolsep{4pt}
\begin{tabularx}{\linewidth}{>{\RaggedRight\arraybackslash}X
>{\RaggedRight\arraybackslash}X
>{\RaggedRight\arraybackslash}X}
\toprule
{\bf Access-control} & {\bf Injected Changes}  & {\bf Systems Applied} \\
\midrule
File permission & allow/block file access & \ucsdcse{}, \opera{}, \sysnet{} \\
\hline
Web server & allow/block URL      & \ucsdcse{}, \opera{}, \sysnet{} \\
access-control & allow/block IP   & \company{}   \\
\hline
App access-control & allow/block pages edit     & Wikipedia  \\
\bottomrule
\end{tabularx}
\caption{
\textbf{Different types of access-control configuration changes injected
  in our experiments.} 
}
\label{tab:injectedchange}
}
\end{table}

Based on the collected configurations, access-control configuration changes are randomly injected to evaluate \acltest{}s. 
The injected changes are generated based on the access-control mechanisms each system is configured with.
Table~\ref{tab:injectedchange} shows the three types of access-control mechanisms used in the systems and the corresponding types of changes injected to them.
Two change types are from previous work~\cite{xiang2019towards} and one change type is added by us for web application permissions. 
For each change type, a number of the change targets (e.g. file, directory, URL) are randomly chosen to apply the change. The number of the change targets is set to be 10\% of all the available resources.

Two sets of experiments are conducted to measure \acltest{}s' effectiveness under different settings: different trimming methods and different ways to generate test inputs.

\subsubsection{Comparing Different Trimming Methods}

We use both strawman and advanced trimming methods as discussed in \S\ref{sec:optimization}
to generate \acltest{}s for different systems.
To better understand the limit of trimming methods alone and avoid the disturbance of historical logs, synthesized requests are used as test inputs.

The advanced trimming works effectively on all systems in terms of detecting impacted requests.
Table~\ref{tab:injectchangeres_trim} shows that \acltest{}s with the advanced trimming can detect all the impacted requests in different systems.
This indicates that the advanced trimming correctly keeps access-control checks when removing unrelated computation and I/O.

The strawman trimming's effectiveness varies on systems.
As Table~\ref{tab:injectchangeres_trim} shows, Wikipedia's \acltest{} with the strawman trimming can detect none of the impacted requests, while \company{}'s \acltest{} with strawman trimming can detect all the impacted requests. For \ucsdcse{}, \sysnet{} and \opera{}, the \acltest{}s can detect 46\%-53\% of the impacted changes. 
The reason is that different systems have different implementations of access-control mechanisms.
Wikipedia's access-control mechanism is implemented by PHP code, which is done mostly in request sub-handlers that are removed by the strawman trimming.
\company{} uses Nginx's access-control, which is implemented out of request sub-handlers and so is not removed by the strawman trimming.
For \ucsdcse{}, \sysnet{} and \opera{}, they use both Nginx/Apache web server's access-control and file permissions checks. 
Nginx/Apache's access checks are kept with the strawman trimming but file permission checks are all removed by the strawman trimming.

\subsubsection{Comparing Different Ways of Generating Test Input}
We evaluate \acltest{}s' effectiveness with test inputs from replaying historical logs and synthesizing requests separately.
Synthesizing requests use the specified source of subjects, objects and actions (e.g., database tables and file directories) to generate test requests.

By synthesizing requests, \acltest{}s detect all the impacted requests, as shown in Table~\ref{tab:injectchangeres_log}.
This illustrates that request synthesis can generate a comprehensive set of test requests.
We admit that this depends on sysadmins' knowledge on the subjects, objects and actions.
From our experience, most of them can be specified as database tables and file directories, and \acltest{}s can automatically extract the possible values for synthesis.

\begin{table}[t!]
{
\centering
\setlength\tabcolsep{2pt}
\begin{tabular}{lccc}
\toprule
{\bf Real-world} & {\bf \# of Total}   &  \multicolumn{2}{c}{\bf \# (\%) of detection } \\
\cline{3-4}
{\bf systems} & {\bf injections}   &  {\bf Strawman trim }  & {\bf Advanced trim}\\
\midrule
Wikipedia  & 930  & 0 (0\%)      & 930 (100\%) \\
\company{} & 4861 & 4861 (100\%)   & 4861 (100\%) \\
\ucsdcse{} & 303  & 142 (47\%) & 303 (100\%) \\
\sysnet{}  & 710  & 329 (46\%) & 710 (100\%) \\
\opera{}   & 178  & 94 (53\%) & 178 (100\%) \\
\bottomrule
\end{tabular}
\caption{
\textbf {Detecting change impacts with \acltest{}s generated by different trimming methods.}}
\label{tab:injectchangeres_trim}
}
\end{table}

\begin{table}[t!]
{
\centering
\setlength\tabcolsep{2pt}
\begin{tabular}{lccc}
\toprule
{\bf Real-world} & {\bf\# of total}   &  \multicolumn{2}{c}{\bf \# (\%) of detection } \\
\cline{3-4}
{\bf systems} & {\bf injections}   &  {\bf Replay logs }  & {\bf Synthesize requests}\\
\midrule
Wikipedia  &930 & 739 (80\%) &  930 (100\%) \\
\company{} &4861& 0 (0\%)    &  4861 (100\%) \\
\ucsdcse{} &303 & 242 (80\%) &  303 (100\%) \\
\sysnet{}  &710 & 106 (15\%) &  710 (100\%) \\
\opera{}   &178 & 19 (11\%)  &  178 (100\%) \\
\bottomrule
\end{tabular}
\caption{
\textbf{Detecting change impacts with different ways of generating test inputs.}}
\label{tab:injectchangeres_log}
}
\end{table}

By replaying historical logs, \acltest{}s can detect up to 80\% of the impacted requests, as shown in Table~\ref{tab:injectchangeres_log}.
\acltestcap{}s work best on Wikipedia and \ucsdcse{}, 
because their logs have a good coverage of possible requests.
For Wikipedia, the logs span 18 years and cover requests to most pages.
For \ucsdcse{}, the logs span 1 year and the web pages are also about classes in the last one year.
For \sysnet{} and \opera{}, the logs fail to cover many obsolete web pages that were never modified in the last five years.
For \company{}, replaying log can detect no change, as the time span is only 1.4 days.
This illustrates that for systems with actively-accessed resources and complete historical logs, sysadmins may use historical logs to generate test requests for \acltest{}s; otherwise, it is better to use synthesized requests to have a better coverage.

\subsection{Performance}

We measured the running time of using \acltest{}s to test the five real systems. 
Synthesizing requests are used to make sure that \acltest{}s have a good coverage of test requests.
The results show that for \company{}, \ucsdcse{}, \sysnet{} and \opera{}, it only takes less than 10 minutes to finish all the tests.
For Wikipedia, it takes a relatively longer time, 1.2 hours. 
The main bottleneck for Wikipedia is the database queries that cannot be trimmed. 

We compared \acltest{}s' performance with different trimming methods: no trimming, strawman trimming and advanced trimming.
No trim means that original programs are used as test programs, which is our baseline.
The strawman trimming does not ensure correct access-control results all the time but can be treated as an upper bound performance that can be achieved by trimming.
The advanced trimming precisely skips I/O and computation unrelated to access-control checks and always ensures correct access-control results.

\begin{figure}[t!]
    \centering
    \includegraphics[width=0.47\textwidth]{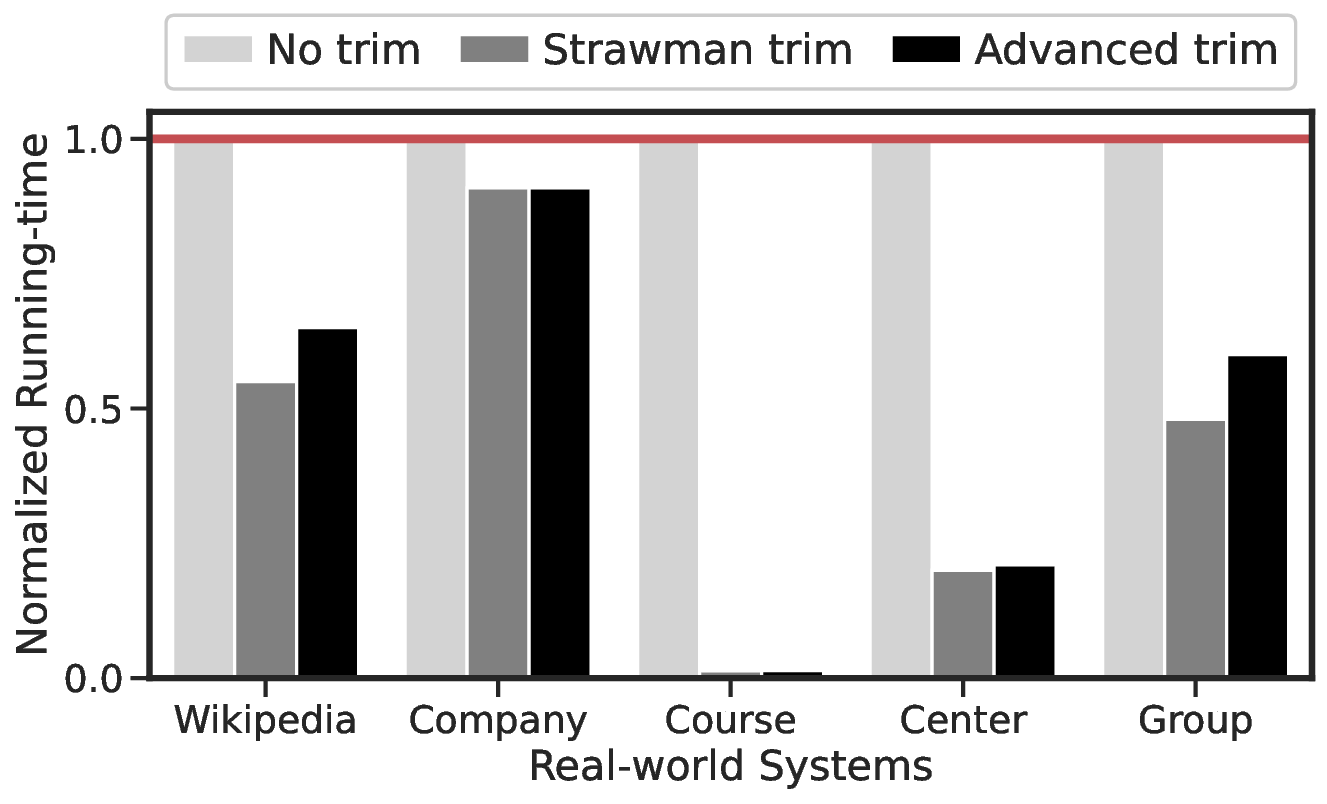}
    \caption{
    \textbf{Normalized test running-time by different trim methods.}
    }
    \label{fig:performance}
\end{figure}

As shown in Figure~\ref{fig:performance}, advanced trim can achieve comparable speedup as strawman trim.
Advanced trim reduced 9.09\%-98.61\% running time compared with no trim, while strawman trim reduced 9.09\%-98.64\%. 
Both advanced and strawman trim work best (98.61\% and 98.64\% reduction) on the \ucsdcse{} system. 
The reason is that \ucsdcse{} has many large files in PDF and MP4 formats. 
It takes a long time to do I/O with no trim, while advanced and strawman trim eliminate the necessity to do I/O.
Both advanced trim and strawman trim have only a small speedup on the \company{} system.
This is because the system is a web proxy, which has a fast request handler only for redirecting requests.

\section{Discussions}
\label{sec:discuss}

The generation and distribution of \acltest{}s can be merged into software vendors' release process of their programs.
When a new version of a program is ready to release,
vendors can use our \acltestgen{} to generate an \acltest{}
and distribute the \acltest{} along with the new program.
To use \acltestgen{}, the developers need to provide inputs for the dynamic analysis.
These are the same inputs for the functional tests of the original programs' access-control, which is necessary for the development process in the first place.
The users, such as sysadmins, can download and setup the \acltest{} along with the original program. 
In the future, when sysadmins make an access-control configuration change, they can run the \acltest{} to test the change impacts.

Our trimming method has several limitations.
First, as it uses dynamic analysis, it does not analyze every path and so can erroneously trim access-control checks.
Although we introduced static analysis to avoid that, the static analysis still does not explore all paths and so can still make mistakes.
This can make the trimmed programs generate wrong access check results and so detect false changes as well miss true changes.
To filter out false changes, we introduced the second-round test with non-trimmed programs. 
Missing true changes is still a limitation, although our evaluation in \S\ref{sec:exp_inject} shows that no true change is missed because of trimming.
Second, trimming cannot eliminate all time-consuming I/Os. 
Some I/Os may be needed for access-control checks to make the decisions, like reading a user table.
Luckily, these I/Os usually do not account for the major execution time.

For test request generation, the choice between using access logs and synthesizing requests can be made based on the specific scenario.
Access logs provide a convenient way to generate test requests, but the test effectiveness depends on the coverage of historical requests. 
With this, sysadmins pay almost no effort and need no knowledge about the subjects, objects and actions.
On the other hand, test request synthesis provides a complete set of test inputs.
But it requires sysadmins to specify the subjects, objects and actions. 
So this may not suit for all sysadmins; however, it at least provides a way for experienced sysadmins to thoroughly test their systems.
Web scanners can also be used to generate test requests for \acltest{}s. Note web
scanners cannot replace \acltest{}s, because it doesn't address the issue of running web applications safely and efficiently with production environment. In addition, from our experiments, web scanners generate a subset of the synthesized requests--web scanners can’t cover customized paths.
\section{Related Work}

\subsection{Access-control testing and verification}
Testing and verification techniques for access-control policies were previously explored mainly by the software engineering community, as a promising approach to detect access-control misconfigurations~\cite{martin2007automated, martin2007fault, pretschner2008model,hu2017verification,fisler2005verification, hughes2008automated, jha2008towards, nanevski2011verification, jayaraman2011automatic}. 
The basic idea is to write or generate test cases that assert on the access results of test requests (i.e., test inputs). With comprehensive test inputs and oracles, misconfigurations will fail the tests.

Prior techniques on access-control testing and verification are based on formal modeling (e.g. using ACML). However, it is non-trivial to encode the actual system configurations using the modeling language, due to the diversity and complexity of access-control implementations~\cite{xiang2019towards}. Furthermore, it takes significant effort to maintain the consistency between the model and system configuration changes, which has been reported as one of the main reasons that impair access-control in real world~\cite{sinclair:10}. Another major obstacle is that the prior approaches require sysadmins to provide oracles or specifications (to check the test results).
With the velocity of access-control changes~\cite{sinclair:10,das2010baaz}, it is untenable to maintain consistency or oracles/specifications.

\acltest{}s address the fundamental limitations of prior access-control testing and verification approaches---it requires neither formal policy models nor oracles/specifications:
1) \acltest{}s take advantage of the original program to test its own access-control configurations to achieve high fidelity;
2) \acltest{}s focus on the impact of access-control changes and flags high-impact changes, without the need for sysadmins to write test oracles or specifications.

\subsection{Access-control misconfiguration detection}

Prior work proposes to detect access-control misconfigurations based on the heuristic that users tend to have similar permissions to similar resources~\cite{das2010baaz, bauer2011detecting, shaikh2017data}.
Those techniques classify users and resources into groups and identify misconfigurations when the same group of users has different permissions to the same group of resources.
Such techniques can only detect limited types of misconfigurations where the heuristic applies.
As noted by Das et al.~\cite{das2010baaz},
``{\it We do not claim that techniques will
find all misconfigurations, as the notion of policy itself is
not defined in most of our deployment settings. Also,
given that access permissions change very organically
over time and several of these changes are linked to adhoc and one-off access requirements, it is very difficult
for an automated system to deduce the exact and complete list of all misconfigurations.}''
Our work presents a complementary approach to focus on detecting changes of dynamic system behaviors instead of the similarity of static permissions.

\subsection{Configuration testing}

A few recent works also propose to use testing techniques to detect misconfigurations~\cite{sun2020testing,xu:16}.
We share the same view that testing is a practical and effective approach for misconfiguration detection;
however, \acltest{}s are fundamentally different.
First, no prior work can be applied to access-control configurations, because they rely on failure symptoms to determine the correctness of configuration values (e.g., crashing behavior, exceptions, and performance degradation). 
However, access-control misconfigurations do not lead to clear failure symptoms.
For the same reason, most of the existing misconfiguration detection techniques for functional correctness and performance cannot address access-control misconfigurations~\cite{zhang:14, tang2015holistic, huang2015confvalley, potharaju2015confseer, santolucito2017synthesizing, baset2017usable, mehta2020rex, chen2020understanding, xiang2020pracextractor}.
Moreover, existing configuration testing techniques are limited to unit-level tests~\cite{xu:16,sun2020testing}. Differently, \acltest{}s focus on system-level tests to reason about end-to-end access-control. This requires addressing the system challenges to run expensive tests efficiently.

\subsection{Execution acceleration}
Existing techniques~\cite{tucek2009efficient,kim2012efficient,tan2017efficient,dou2019shortcut} for accelerating execution aim to reduce the number of requests to execute. 
They aggregate similar requests, execute them in one round for the most part and only split the executions when there are divergences on control or data flow.
\acltest{}s take an orthogonal approach. It minimizes the execution time of each individual request by trimming the original server program.
We observe that server programs typically perform access-control checks at the beginning of a request handling; therefore, we can skip the costly operations after the access-control checks.
To make \acltest{}s easy to deploy, our approach does not introduce any new dependency or record-and-replay systems as prior approaches. 
Combining our trimming and previous deduplication approaches may further reduce the testing  time.
\section{Conclusion}

This paper presents \acltest{}s, a new type of mini test environment for testing access control configuration changes in web applications.
\acltestcap{}s can detect what requests are impacted by access control configuration changes and warn sysadmins to validate if the impacts are intended.
This can help sysadmins detect vulnerabilities before they are exploited by attackers.
To help developers build \acltest{}s, we present a general technique to transform a target program into an \acltest{}.
Our evaluation on real-world Docker images shows that \acltest{}s detect 168 new vulnerabilities from 72 images. 
So far 54 of these vulnerabilities have been confirmed and 44 of them have been fixed by image maintainers.
Our evaluation on five real-world deployed systems shows that \acltest{}s can effectively and efficiently detect all the change impacts.

As many other tools, \acltest{} is far from perfect. As for future research, there are several promising directions to be explored. 
First, test requests generation is a major challenge for web application testing. Future work may need to explore how to generate request parameters with a high coverage. 
Second, automatic triage of access-control behavior changes is a new problem to be explored. 
\acltestcap{} only uses several simple rules to triage a limited number of changes;
however, more advanced triage algorithms can improve the coverage.
Third, \acltest{} only targets on web applications. 
Testing access-control changes in other types of systems is still an important but unsolved problem.

{
\bibliographystyle{plain}
\bibliography{ref}

\begin{thebibliography}{10}

\bibitem{vul_git}
{Acunetix}.
\newblock Git repository found.
\newblock \url{https://www.acunetix.com/vulnerabilities/web/git-repository-found/}, 2021.

\bibitem{vul_htaccess}
{Acunetix}.
\newblock .htaccess file readable.
\newblock \url{https://www.acunetix.com/vulnerabilities/web/htaccess-file-readable/}, 2021.

\bibitem{adrion1982validation}
W~Richards Adrion, Martha~A Branstad, and John~C Cherniavsky.
\newblock Validation, verification, and testing of computer software.
\newblock {\em ACM Computing Surveys (CSUR)}, 14(2):159--192, 1982.

\bibitem{testinproduction2}
{Ann Marie}.
\newblock Don't test in production? test in production!
\newblock \url{https://opensource.com/article/19/5/dont-test-production}, 2019.

\bibitem{apachemanual}
{Apache Foundation}.
\newblock Httpd 2.4 documentation.
\newblock \url{https://httpd.apache.org/docs/2.4/howto/access.html}, 2021.

\bibitem{armstrong2013linkbench}
Timothy~G Armstrong, Vamsi Ponnekanti, Dhruba Borthakur, and Mark Callaghan.
\newblock Linkbench: a database benchmark based on the facebook social graph.
\newblock In {\em Proceedings of the 2013 ACM SIGMOD International Conference on Management of Data}, pages 1185--1196, 2013.

\bibitem{baset2017usable}
Salman Baset, Sahil Suneja, Nilton Bila, Ozan Tuncer, and Canturk Isci.
\newblock Usable declarative configuration specification and validation for applications, systems, and cloud.
\newblock In {\em Proceedings of the 18th ACM/IFIP/USENIX Middleware Conference: Industrial Track}, pages 29--35, 2017.

\bibitem{bauer2011detecting}
Lujo Bauer, Scott Garriss, and Michael~K Reiter.
\newblock Detecting and resolving policy misconfigurations in access-control systems.
\newblock {\em ACM Transactions on Information and System Security (TISSEC)}, 14(1):2, 2011.

\bibitem{BrAPI}
{BrAPI}.
\newblock Brapi.
\newblock \url{https://wiki.brapi.org/index.php/BrAPI} is the deployed website of the image \url{https://hub.docker.com/r/brapicoordinatorselby/brapi-wiki} we detected vulnerability, 2019.

\bibitem{cao2020characterizing}
Zhichao Cao, Siying Dong, Sagar Vemuri, and David~HC Du.
\newblock Characterizing, modeling, and benchmarking rocksdb key-value workloads at facebook.
\newblock In {\em 18th USENIX Conference on File and Storage Technologies (FAST 20)}, pages 209--223, 2020.

\bibitem{chen2020understanding}
Qingrong Chen, Teng Wang, Owolabi Legunsen, Shanshan Li, and Tianyin Xu.
\newblock Understanding and discovering software configuration dependencies in cloud and datacenter systems.
\newblock In {\em Proceedings of the 28th ACM Joint Meeting on European Software Engineering Conference and Symposium on the Foundations of Software Engineering}, pages 362--374, 2020.

\bibitem{cheng2021test}
Runxiang Cheng, Lingming Zhang, Darko Marinov, and Tianyin Xu.
\newblock Test-case prioritization for configuration testing.
\newblock In {\em Proceedings of the 30th ACM SIGSOFT International Symposium on Software Testing and Analysis}, pages 452--465, 2021.

\bibitem{vsftpdmanual}
{Chris Evans}.
\newblock Vsftpd man page.
\newblock \url{https://linux.die.net/man/5/vsftpd.conf}, 2021.

\bibitem{testinproduction}
{Cindy Sridharan}.
\newblock Testing in production, the safe way.
\newblock \url{https://copyconstruct.medium.com/testing-in-production-the-safe-way-18ca102d0ef1}, 2018.

\bibitem{cfg-diff}
{conf-test}.
\newblock Cfg-diff graphs.
\newblock \url{https://github.com/conf-test/acl-test/tree/main/cfg-diff}, 2021.

\bibitem{apachetriminput}
{conf-test}.
\newblock Httpd trim input.
\newblock \url{https://github.com/conf-test/acl-test/blob/main/httpd/trim_run.sh}, 2021.

\bibitem{nginxtriminput}
{conf-test}.
\newblock Nginx trim input.
\newblock \url{https://github.com/conf-test/acl-test/blob/main/nginx/trim_run.sh}, 2021.

\bibitem{postgrestriminput}
{conf-test}.
\newblock Postgresql trim input.
\newblock \url{https://github.com/conf-test/acl-test/blob/main/postgres/trim_run.sh}, 2021.

\bibitem{linuxserver_report}
{conf-test}.
\newblock Sensitive files are exposed.
\newblock \url{https://github.com/linuxserver/docker-dokuwiki/issues/33}, 2021.

\bibitem{vsftpdtriminput}
{conf-test}.
\newblock Vsftpd trim input.
\newblock \url{https://github.com/conf-test/acl-test/blob/main/vsftpd/trim_run.sh}, 2021.

\bibitem{full_list_vul}
{conf-test}.
\newblock The full list of our detected vulnerabilities (anonymous access).
\newblock \url{https://docs.google.com/spreadsheets/d/18_oaaoEawPJ-GNajqEL7t43tiKQ2iizZ9J6URCZMWc0}, 2022.

\bibitem{gunPermit}
{Dan Walters}.
\newblock How did confidential gun permit data get leaked?
\newblock \url{https://calmatters.org/commentary/2022/12/how-did-confidential-gun-permit-data-get-leaked/}, 2022.

\bibitem{das2010baaz}
Tathagata Das, Ranjita Bhagwan, and Prasad Naldurg.
\newblock {Baaz: A System for Detecting Access Control Misconfigurations}.
\newblock In {\em Proceedings of the 19th USENIX Security Symposium (USENIX Security'10)}, August 2010.

\bibitem{docker-file}
{Docker}.
\newblock Dockerfile reference.
\newblock \url{https://docs.docker.com/engine/reference/builder/}, 2021.

\bibitem{dockerhub}
{Docker}.
\newblock Dockerhub.
\newblock \url{https://hub.docker.com/search?q=&type=image}, 2021.

\bibitem{overlaynetwork}
{Docker}.
\newblock Overlay networks.
\newblock \url{https://docs.docker.com/network/overlay/}, 2021.

\bibitem{dokuwiki}
{dokuwiki}.
\newblock Dokuwiki.
\newblock \url{https://www.dokuwiki.org/dokuwiki}, 2021.

\bibitem{dou2019shortcut}
Xianzheng Dou, Peter~M Chen, and Jason Flinn.
\newblock Shortcut: accelerating mostly-deterministic code regions.
\newblock In {\em Proceedings of the 27th ACM Symposium on Operating Systems Principles}, pages 570--585, 2019.

\bibitem{drupal}
{Drupal}.
\newblock Drupal - open source cms.
\newblock \url{https://www.drupal.org/}, 2020.

\bibitem{drupalmodulepage}
{Drupal}.
\newblock Drupal modules.
\newblock \url{https://www.drupal.org/project/project_module}, 2021.

\bibitem{duvall2007continuous}
Paul~M Duvall, Steve Matyas, and Andrew Glover.
\newblock {\em Continuous integration: improving software quality and reducing risk}.
\newblock Pearson Education, 2007.

\bibitem{EliasHolzmann_fp}
{EliasHolzmann}.
\newblock Sensitive data exposed to public access.
\newblock \url{https://github.com/EliasHolzmann/mediawiki-docker/issues/1}, 2021.

\bibitem{fetterly2011tidyfs}
Dennis Fetterly, Maya Haridasan, Michael Isard, and Swaminathan Sundararaman.
\newblock Tidyfs: A simple and small distributed file system.
\newblock In {\em USENIX annual technical conference}, pages 34--34, 2011.

\bibitem{fisler2005verification}
Kathi Fisler, Shriram Krishnamurthi, Leo~A Meyerovich, and Michael~Carl Tschantz.
\newblock {Verification and Change-Impact Analysis of Access-Control Policies}.
\newblock In {\em Proceedings of the 27th International Conference on Software Engineering (ICSE'05)}, May 2005.

\bibitem{vul_vendor}
{hackerone}.
\newblock Potential server misconfiguration leads to disclosure of vendor/ directory.
\newblock \url{https://hackerone.com/reports/271391}, 2017.

\bibitem{vul_composer}
{hackerone}.
\newblock Development configuration file.
\newblock \url{https://hackerone.com/reports/231267}, 2018.

\bibitem{vul_phpunit}
{hackerone}.
\newblock Sensitive data disclosure via exposed phpunit file.
\newblock \url{https://hackerone.com/reports/543775}, 2020.

\bibitem{harter2014analysis}
Tyler Harter, Dhruba Borthakur, Siying Dong, Amitanand Aiyer, Liyin Tang, Andrea~C Arpaci-Dusseau, and Remzi~H Arpaci-Dusseau.
\newblock Analysis of hdfs under hbase: A facebook messages case study.
\newblock In {\em 12th USENIX Conference on File and Storage Technologies (FAST 14)}, pages 199--212, 2014.

\bibitem{hu2017verification}
Vincent~C Hu, Rick Kuhn, and Dylan Yaga.
\newblock Verification and test methods for access control policies/models.
\newblock {\em NIST Special Publication}, 800:192, 2017.

\bibitem{huang1975approach}
JC~Huang.
\newblock An approach to program testing.
\newblock {\em ACM Computing Surveys (CSUR)}, 7(3):113--128, 1975.

\bibitem{huang2015confvalley}
Peng Huang, William~J Bolosky, Abhishek Singh, and Yuanyuan Zhou.
\newblock Confvalley: A systematic configuration validation framework for cloud services.
\newblock In {\em Proceedings of the Tenth European Conference on Computer Systems}, pages 1--16, 2015.

\bibitem{hughes2008automated}
Graham Hughes and Tevfik Bultan.
\newblock Automated verification of access control policies using a sat solver.
\newblock {\em International journal on software tools for technology transfer}, 10(6):503--520, 2008.

\bibitem{humble2010continuous}
Jez Humble and David Farley.
\newblock {\em Continuous delivery: reliable software releases through build, test, and deployment automation}.
\newblock Pearson Education, 2010.

\bibitem{jayaraman2011automatic}
Karthick Jayaraman, Vijay Ganesh, Mahesh Tripunitara, Martin Rinard, and Steve Chapin.
\newblock Automatic error finding in access-control policies.
\newblock In {\em Proceedings of the 18th ACM conference on Computer and communications security}, pages 163--174, 2011.

\bibitem{jha2008towards}
Somesh Jha, Ninghui Li, Mahesh Tripunitara, Qihua Wang, and William Winsborough.
\newblock Towards formal verification of role-based access control policies.
\newblock {\em IEEE Transactions on Dependable and Secure Computing}, 5(4):242--255, 2008.

\bibitem{immediateHealth}
{Jill McKeon}.
\newblock Inmediata health reaches \$1.13m settlement after 2019 data breach.
\newblock \url{https://healthitsecurity.com/news/inmediata-health-reaches-1.13m-settlement-after-2019-data-breach}, 2022.

\bibitem{joomla}
{joomla}.
\newblock Joomla content management system (cms).
\newblock \url{https://www.joomla.org/}, 2021.

\bibitem{kim2012efficient}
Taesoo Kim, Ramesh Chandra, and Nickolai Zeldovich.
\newblock Efficient patch-based auditing for web application vulnerabilities.
\newblock In {\em 10th USENIX Symposium on Operating Systems Design and Implementation (OSDI 12)}, pages 193--206, 2012.

\bibitem{lampson:09}
Butler Lampson.
\newblock {Usable Security: How to Get It}.
\newblock {\em Communications of the ACM}, 52(11):25--27, November 2009.

\bibitem{lattner2004llvm}
Chris Lattner and Vikram Adve.
\newblock Llvm: A compilation framework for lifelong program analysis \& transformation.
\newblock In {\em International Symposium on Code Generation and Optimization, 2004. CGO 2004.}, pages 75--86. IEEE, 2004.

\bibitem{anyip}
{Linux}.
\newblock ipv6: Implement any-ip support for ipv6.
\newblock \url{https://git.kernel.org/pub/scm/linux/kernel/git/torvalds/linux.git/commit/?id=ab79ad14a2d51e95f0ac3cef7cd116a57089ba828}, 2010.

\bibitem{overlayfs}
{Linux}.
\newblock Overlay filesystem.
\newblock \url{https://www.kernel.org/doc/html/latest/filesystems/overlayfs.html}, 2021.

\bibitem{linuxserver_fp}
{linuxserver}.
\newblock Sensitive files are exposed.
\newblock \url{https://github.com/linuxserver/docker-dokuwiki/issues/33#issuecomment-820429021}, 2021.

\bibitem{linuxserver}
{linuxserver.io}.
\newblock linuxserver/dokuwiki.
\newblock \url{https://hub.docker.com/r/linuxserver/dokuwiki}, 2022.

\bibitem{llvm-analysis}
{LLVM}.
\newblock Llvm analysis passes.
\newblock \url{https://llvm.org/docs/Passes.html#analysis-passes}, 2021.

\bibitem{llvm-transform}
{LLVM}.
\newblock Llvm transform passes.
\newblock \url{https://llvm.org/docs/Passes.html#transform-passes}, 2021.

\bibitem{lorch2003first}
Markus Lorch, Seth Proctor, Rebekah Lepro, Dennis Kafura, and Sumit Shah.
\newblock First experiences using xacml for access control in distributed systems.
\newblock In {\em Proceedings of the 2003 ACM workshop on XML security}, pages 25--37, 2003.

\bibitem{martin2007automated}
Evan Martin and Tao Xie.
\newblock {Automated Test Generation for Access Control Policies via Change-Impact Analysis}.
\newblock In {\em Proceedings of the 3rd International Workshop on Software Engineering for Secure Systems}, 2007.

\bibitem{martin2007fault}
Evan Martin and Tao Xie.
\newblock A fault model and mutation testing of access control policies.
\newblock In {\em Proceedings of the 16th international conference on World Wide Web}, pages 667--676, 2007.

\bibitem{mehta2020rex}
Sonu Mehta, Ranjita Bhagwan, Rahul Kumar, Chetan Bansal, Chandra Maddila, B~Ashok, Sumit Asthana, Christian Bird, and Aditya Kumar.
\newblock Rex: Preventing bugs and misconfiguration in large services using correlated change analysis.
\newblock In {\em 17th USENIX Symposium on Networked Systems Design and Implementation (NSDI 20)}, pages 435--448, 2020.

\bibitem{nanevski2011verification}
Aleksandar Nanevski, Anindya Banerjee, and Deepak Garg.
\newblock Verification of information flow and access control policies with dependent types.
\newblock In {\em 2011 IEEE Symposium on Security and Privacy}, pages 165--179. IEEE, 2011.

\bibitem{nginxmanual}
{Nginx}.
\newblock Nginx admin guide.
\newblock \url{https://docs.nginx.com/nginx/admin-guide/security-controls/configuring-http-basic-authentication/}, 2021.

\bibitem{PHP-Parser}
{Nikita Popov}.
\newblock A php parser written in php.
\newblock \url{https://github.com/nikic/php-parser}, 2023.

\bibitem{nishtala2013scaling}
Rajesh Nishtala, Hans Fugal, Steven Grimm, Marc Kwiatkowski, Herman Lee, Harry~C Li, Ryan McElroy, Mike Paleczny, Daniel Peek, Paul Saab, et~al.
\newblock Scaling memcache at facebook.
\newblock In {\em 10th USENIX Symposium on Networked Systems Design and Implementation (NSDI 13)}, pages 385--398, 2013.

\bibitem{nlpub.ru}
{nlpub.ru}.
\newblock nlpub.ru.
\newblock \url{https://nlpub.ru/} is the deployed website of the image \url{https://hub.docker.com/r/nlpub/mediawiki} we detected vulnerability, 2019.

\bibitem{postgresmanual}
{PostgreSQL}.
\newblock Postgresql documentation.
\newblock \url{https://www.postgresql.org/docs/current/ddl-priv.html}, 2021.

\bibitem{potharaju2015confseer}
Rahul Potharaju, Joseph Chan, Luhui Hu, Cristina Nita-Rotaru, Mingshi Wang, Liyuan Zhang, and Navendu Jain.
\newblock Confseer: leveraging customer support knowledge bases for automated misconfiguration detection.
\newblock {\em Proceedings of the VLDB Endowment}, 8(12):1828--1839, 2015.

\bibitem{pretschner2008model}
Alexander Pretschner, Tejeddine Mouelhi, and Yves Le~Traon.
\newblock Model-based tests for access control policies.
\newblock In {\em 2008 1st International Conference on Software Testing, Verification, and Validation}, pages 338--347. IEEE, 2008.

\bibitem{rodeh2013btrfs}
Ohad Rodeh, Josef Bacik, and Chris Mason.
\newblock Btrfs: The linux b-tree filesystem.
\newblock {\em ACM Transactions on Storage (TOS)}, 9(3):1--32, 2013.

\bibitem{santolucito2017synthesizing}
Mark Santolucito, Ennan Zhai, Rahul Dhodapkar, Aaron Shim, and Ruzica Piskac.
\newblock Synthesizing configuration file specifications with association rule learning.
\newblock {\em Proceedings of the ACM on Programming Languages}, 1(OOPSLA):1--20, 2017.

\bibitem{servarr}
{servarr}.
\newblock servarr/mediawiki:1.0.7.
\newblock \url{https://hub.docker.com/layers/servarr/mediawiki/1.0.7/images/sha256-610718f59d235a7d7aab07bc59c1a86e1ede267361c38159942e705a13b7fa4b?context=explore}, 2022.

\bibitem{shahin2017continuous}
Mojtaba Shahin, Muhammad~Ali Babar, and Liming Zhu.
\newblock Continuous integration, delivery and deployment: a systematic review on approaches, tools, challenges and practices.
\newblock {\em IEEE access}, 5:3909--3943, 2017.

\bibitem{shaikh2017data}
Riaz~Ahmed Shaikh, Kamel Adi, and Luigi Logrippo.
\newblock {A Data Classification Method for Inconsistency and Incompleteness Detection in Access Control Policy Sets}.
\newblock {\em International Journal of Information Security}, 16(1):91--113, 2017.

\bibitem{sinclair:10}
Sara Sinclair and Sean~W. Smith.
\newblock {What's Wrong with Access Control in the Real World?}
\newblock {\em IEEE Security \& Privacy}, 8(4):74--77, July 2010.

\bibitem{sun2020testing}
Xudong Sun, Runxiang Cheng, Jianyan Chen, Elaine Ang, Owolabi Legunsen, and Tianyin Xu.
\newblock Testing configuration changes in context to prevent production failures.
\newblock In {\em 14th USENIX Symposium on Operating Systems Design and Implementation (OSDI 20)}, pages 735--751, 2020.

\bibitem{tan2017efficient}
Cheng Tan, Lingfan Yu, Joshua~B Leners, and Michael Walfish.
\newblock The efficient server audit problem, deduplicated re-execution, and the web.
\newblock In {\em Proceedings of the 26th Symposium on Operating Systems Principles}, pages 546--564, 2017.

\bibitem{tang2015holistic}
Chunqiang Tang, Thawan Kooburat, Pradeep Venkatachalam, Akshay Chander, Zhe Wen, Aravind Narayanan, Patrick Dowell, and Robert Karl.
\newblock Holistic configuration management at facebook.
\newblock In {\em Proceedings of the 25th Symposium on Operating Systems Principles}, pages 328--343, 2015.

\bibitem{wordpressplugin}
{Tech Times }.
\newblock Wordpress data breach affects 100,000 exposed websites after using responsive menu plugin.
\newblock \url{https://www.techtimes.com/articles/257016/20210212/wordpress-data-breach-affects-100-000-exposed-websites-using-responsive.htm}, 2021.

\bibitem{terrace2009object}
Jeff Terrace and Michael~J Freedman.
\newblock Object storage on craq: High-throughput chain replication for read-mostly workloads.
\newblock In {\em USENIX Annual Technical Conference}, number June, pages 1--16. San Diego, CA, 2009.

\bibitem{tkw1536_fp}
{tkw1536}.
\newblock Risky data exposed to public access.
\newblock \url{https://github.com/tkw1536/docker-dokuwiki/issues/1}, 2021.

\bibitem{trumpCampaign}
{Tom Spring}.
\newblock Trump campaign website left open to email server hijack.
\newblock \url{https://threatpost.com/trump-campaign-website-allowed-email-hijack/149278/}, 2019.

\bibitem{tucek2009efficient}
Joseph Tucek, Weiwei Xiong, and Yuanyuan Zhou.
\newblock Efficient online validation with delta execution.
\newblock In {\em Proceedings of the 14th international conference on Architectural support for programming languages and operating systems}, pages 193--204, 2009.

\bibitem{vulfocus}
{vulfocus}.
\newblock vulfocus/drupal-cve\_2019\_6340.
\newblock \url{https://hub.docker.com/r/vulfocus/drupal-cve\_2019\_6340}, 2019.

\bibitem{wang2023test}
Shuai Wang, Xinyu Lian, Darko Marinov, and Tianyin Xu.
\newblock Test selection for unified regression testing.
\newblock In {\em 2023 IEEE/ACM 45th International Conference on Software Engineering (ICSE)}, pages 1687--1699. IEEE, 2023.

\bibitem{west:08}
Ryan West.
\newblock {The Psychology of Security}.
\newblock {\em Communications of the ACM}, 51(4):34--40, April 2008.

\bibitem{cowiki}
{wikimedia}.
\newblock cowikimedia dump progress).
\newblock \url{https://dumps.wikimedia.org/cowikimedia/20221201/}, 2022.

\bibitem{mediawikiextensionpage}
{WikiMedia Foundation }.
\newblock Extensions let you customize how mediawiki looks and works.
\newblock \url{https://www.mediawiki.org/wiki/Manual:Extensions}, 2021.

\bibitem{mediawiki}
{WikiMedia Foundation}.
\newblock Mediawiki is a collaboration and documentation platform brought to you by a vibrant community.
\newblock \url{https://www.mediawiki.org/wiki/MediaWiki}, 2020.

\bibitem{wordpresspluginpage}
{Wordpress }.
\newblock Extend your wordpress experience with 58,390 plugins.
\newblock \url{https://wordpress.org/plugins/}, 2021.

\bibitem{wordpress}
{wordpress}.
\newblock Wordpress.com: Create a free website or blog.
\newblock \url{https://wordpress.com/read}, 2021.

\bibitem{xdebug}
{Xdebug}.
\newblock Code coverage analysis.
\newblock \url{https://xdebug.org/docs/code_coverage}, 2023.

\bibitem{xiang2020pracextractor}
Chengcheng Xiang, Haochen Huang, Andrew Yoo, Yuanyuan Zhou, and Shankar Pasupathy.
\newblock Pracextractor: Extracting configuration good practices from manuals to detect server misconfigurations.
\newblock In {\em 2020 USENIX Annual Technical Conference (USENIX ATC 20)}, pages 265--280, 2020.

\bibitem{xiang2019towards}
Chengcheng Xiang, Yudong Wu, Bingyu Shen, Mingyao Shen, Haochen Huang, Tianyin Xu, Yuanyuan Zhou, Cindy Moore, Xinxin Jin, and Tianwei Sheng.
\newblock Towards continuous access control validation and forensics.
\newblock In {\em Proceedings of the 2019 ACM SIGSAC Conference on Computer and Communications Security}, pages 113--129, 2019.

\bibitem{xu2018automated}
Dianxiang Xu, Roshan Shrestha, and Ning Shen.
\newblock Automated coverage-based testing of xacml policies.
\newblock In {\em Proceedings of the 23nd ACM on Symposium on Access Control Models and Technologies}, pages 3--14, 2018.

\bibitem{xu2020automated}
Dianxiang Xu, Roshan Shrestha, and Ning Shen.
\newblock Automated strong mutation testing of xacml policies.
\newblock In {\em Proceedings of the 25th ACM Symposium on Access Control Models and Technologies}, pages 105--116, 2020.

\bibitem{xu:16}
Tianyin Xu, Xinxin Jin, Peng Huang, Yuanyuan Zhou, Shan Lu, Long Jin, and Shankar Pasupathy.
\newblock Early detection of configuration errors to reduce failure damage.
\newblock In {\em Proceedings of the 12th USENIX Conference on Operating Systems Design and Implementation (OSDI'16)}, November 2016.

\bibitem{xu:chi:17}
Tianyin Xu, Han~Min Naing, Le~Lu, and Yuanyuan Zhou.
\newblock {How Do System Administrators Resolve Access-Denied Issues in the Real World?}
\newblock In {\em Proceedings of the 35th Annual CHI Conference on Human Factors in Computing Systems (CHI'17)}, May 2017.

\bibitem{zhang:14}
Jiaqi Zhang, Lakshmi Renganarayana, Xiaolan Zhang, Niyu Ge, Vasanth Bala, Tianyin Xu, and Yuanyuan Zhou.
\newblock {EnCore: Exploiting System Environment and Correlation Information for Misconfiguration Detection}.
\newblock In {\em Proceedings of the 19th International Conference on Architecture Support for Programming Languages and Operating Systems (ASPLOS'14)}, March 2014.

\end{thebibliography}
}
\clearpage
\appendix

\section{Algorithms}

Algorithm~\ref{alg:cfg-diff} consists of two major steps. Firstly, it colors a pair of $\mathsf{cfg}_{\mathsf{allow}}$ and $\mathsf{cfg}_{\mathsf{deny}}$ to identify the diverging nodes as $\mathsf{accCandidates}$. 
Secondly, it identifies true $\mathsf{acc}$ from $\mathsf{accCandidates}$. $\mathsf{accCandidates}$ may includes $\mathsf{non-acc}$ nodes for which the divergence comes from non-deterministic execution of the program.
To filter out these nodes, our algorithm only treats the node with the largest divergence (most number of nodes in the diverged branches) as the final ACC node. This relies on an assumption that access-control decisions cause larger control flow divergence than non-deterministic execution. This assumption holds well throughout our experiments. 

In our implementation of CFG-diff, we also need to handle multiple threads. Server programs usually spawn multiple threads to handle different requests. 
As a results, multiple dynamic CFG will be generated for each run, and we need to decide which CFG to feed into CFG-diff.
We take two steps to handle the challenge. 
First, for our trimming analysis (not the final \acltest{}s), we run requests sequentially so that only one thread is handling request.
Second, we find out the request-handling thread's CFG by searching for the CFG with the most nodes.
Then we use the request-handling thread's CFG for CFG-diff.

\begin{algorithm}[]
\SetAlgoLined
\DontPrintSemicolon
\SetKwProg{Pn}{Function}{:}{\KwRet}
\SetKwFunction{FMain}{diffCFG}
\Pn{\FMain{
    $\mathsf{cfg}_{\mathsf{allow}}$, 
    $\mathsf{cfg}_{\mathsf{deny}}$ } 
    }{
    $\mathsf{cfg}_{\mathsf{diff}}=\mathsf{merge}(\mathsf{cfg}_{\mathsf{allow}}, \mathsf{cfg}_{\mathsf{deny}})$\;
     \For{$\mathsf{node} \in \mathsf{cfg}_{\mathsf{diff}}$}{
      \uIf{$\mathsf{node} \in \mathsf{cfg}_{\mathsf{allow}}$ \textup{\textbf{and}} $\mathsf{node} \in \mathsf{cfg}_{\mathsf{deny}}$}
      {
        $\mathsf{node.color}$(``green-red-mixed'')\;
      }
      \uElseIf{$\mathsf{node} \in \mathsf{cfg}_{\mathsf{allow}}$}
      {
        $\mathsf{node.color}$(``green'')\;
      }
      \Else
      {
        $\mathsf{node.color}$(``red'')\;
      }
     }
     \Return{$\mathsf{cfg}_{\mathsf{diff}}$}
 }
 \SetKwFunction{FMain}{findFinalACCInCFG}
\Pn{
    \FMain{
        $\mathsf{cfg}_{\mathsf{allow}}$, 
        $\mathsf{cfg}_{\mathsf{deny}}$
        } 
    }{
    $\mathsf{cfg}_{\mathsf{diff}}=\mathsf{diffCFG}(\mathsf{cfg}_{\mathsf{allow}}, \mathsf{cfg}_{\mathsf{deny}})$\;
     $\mathsf{accCandidates} = \emptyset{}$\;
     \For{$\mathsf{node} \in \mathsf{cfg}_{\mathsf{diff}}$}{
     \uIf{ $\mathsf{node.color()}==\textup{``green-red-mixed''}$ \textup{\textbf{and}}
            $\mathsf{node.hasGreenChild()}$ \textup{\textbf{and}}
            $\mathsf{node.hasRedChild()}$ 
            }
     {
     $\mathsf{accCandidates.add(node)}$
     }
     }
     $\mathsf{acc} = \mathsf{node WithLargestDiverge(accCandidates)}$\;
     \Return{$\mathsf{acc}$, $\mathsf{cfg}_{\mathsf{diff}}$}
 }
  \caption{CFG-diff to find the final ACC in a pair of allowed and denied CFG}
 \label{alg:cfg-diff}
 
\end{algorithm}

\begin{algorithm}[t]
\begin{minipage}{0.45\textwidth}
\SetAlgoLined
\DontPrintSemicolon
\SetKwProg{Pn}{Function}{:}{\KwRet}
\SetKwFunction{FMain}{forwardAnalysis}
\Pn{\FMain{$\mathsf{acc}$, $\mathsf{cfg}$} }{
$\mathsf{accToAdd=\{acc\}}$\;
\For{$\mathsf{function} \in \mathsf{cfg.calledAfter(acc)}$}{
    \For{$\mathsf{statement} \in \mathsf{function}$ }
    {
        $\mathsf{cfg.staticExpand(function)}$\;
        \If{$\mathsf{isAcc\footnote{$\mathsf{isAcc}$ is implemented by matching function names and return types with the acc functions returned by Algorithm~\ref{alg:cfg-diff}'s $\mathsf{findFinalACCInCFG}$.}(statement)}$}
        {
            $\mathsf{accToAdd.add(statement)}$\;
        }
    }
}
\Return{$\mathsf{accToAdd, cfg}$}
}
\SetKwFunction{FMain}{backwardAnalysis}
\Pn{\FMain{$\mathsf{acc}$, $\mathsf{cfg}$} }{
$\mathsf{accToDel}=\emptyset{}$\;
\For{$\mathsf{function} \in \mathsf{cfg.calledBefore(acc)}$}{
    \For{$\mathsf{statement} \in \mathsf{function}$ }
    {
        $\mathsf{cfg.staticExpand(function)}$\;
        \If{$\mathsf{isAcc(statement)}$}
        {
            $\mathsf{accToDel.add(acc)}$\;
        }
    }
}
\Return{$\mathsf{accToDel, cfg}$}
}
\SetKwFunction{FMain}{findFinalACCs}
\Pn{\FMain{$\mathsf{accInCFGs}$
\footnote{$\mathsf{accInCFGs}$ is the results of Algorithm~\ref{alg:cfg-diff}'s $\mathsf{findFinalACCInCFG}$.}}}
{
$\mathsf{candidates} = \emptyset{}$\;
\For{$\mathsf{acc}$, 
    $\mathsf{cfg}$ 
    $\in \mathsf{accInCFGs}$
    }{
        $\mathsf{accToAdd}, \mathsf{cfg'} =\mathsf{forwardAnalysis(acc}, \mathsf{cfg})$\;
        $\mathsf{candidates} = \mathsf{candidates} \cup \mathsf{accToAdd}$ \;
        $\mathsf{accInCFGs} = \mathsf{accInCFGs} \cup \mathsf{accToAdd}\times \mathsf{cfg'}$
}
\For{$\mathsf{acc}$, 
    $\mathsf{cfg}$ 
    $\in \mathsf{accInCFGs}$
    }{
        $\mathsf{accToDel}, \mathsf{cfg'} =\mathsf{backwardAnalysis(acc}, \mathsf{cfg})$\;
        $\mathsf{candidates} = \mathsf{candidates} - \mathsf{accToDel}$\;
        $\mathsf{accInCFGs} = \mathsf{accInCFGs} \cup \mathsf{accToDel}\times \mathsf{cfg'}$
}
\Return{$\mathsf{candidates}$}
}
 \caption{Static analysis to find the final ACCs}
 \label{alg:forward_backward}
\end{minipage}
\end{algorithm}
\end{document}